\documentclass[12pt,preprint]{aastex}

\shorttitle{HST/NICMOS observations  of Mon R2}
\shortauthors{Andersen et al.}

\received{2006 March 3}
\begin{document}


\title{NICMOS/HST Observations of the Embedded Cluster Associated with Mon R2: Constraining the sub--stellar Initial Mass Function}

\author{M. Andersen, M. R. Meyer, B. Oppenheimer}
\affil{Steward Observatory, University of Arizona, 933 North Cherry Avenue, Tucson, AZ 85721}
\email{mandersen@as.arizona.edu,mmeyer@as.arizona.edu}

\author{C. Dougados}
\affil{Laboratoire d'Astrophysique de Grenoble, France}
\author{ J. Carpenter}
\affil{Department of astronomy, California Institute of Technology, Pasadena, CA91125}

\begin{abstract}
 We have analyzed HST/NICMOS2 F110W, F160W, F165M, and F207M band  images covering the  central 1\mbox{$^\prime$}$\times$1\mbox{$^\prime$} region of the cluster associated with Mon R2 in order to constrain the Initial Mass Function (IMF)  down to  20 M$_\mathrm{jup}$. 
The flux ratio between the F165M and F160W bands  was used to measure the strength of the water band absorption feature and select a sample of 12 out of the total sample of  181 objects that have effective temperatures between 2700 K and 3300 K. 
These objects are placed in the HR diagram together with sources observed by \citet{carpenter} to estimate an  age of $\sim$1 Myr for the low mass cluster population.
By constructing extinction limited samples, we are able to constrain the IMF and the fraction of stars with a circumstellar disk in a sample that is 90\%\ complete for both high and low mass objects. 
For stars with estimated masses between 0.1 M$_\odot$ and 1.0 M$_\odot$ for a 1 Myr population with A$_\mathrm{V} \le 19$ mag, we find that $27\pm9\%$  have a near--infrared excess indicative of a circumstellar disk. 
The derived  fraction is similar to, or slightly lower than, the fraction found in other  star forming regions of comparable age.
We constrain the number of stars in the mass interval 0.08--1.0 M$_\odot$ to the number of objects in the mass interval 0.02--0.08 M$_\odot$ by forming the ratio, $R^{**}$=N(0.08--1 M$_\odot$)/N(0.02--0.08 M$_\odot$) for objects in an extinction limited sample complete for A$_\mathrm{V}\le 7$ mag. 
The ratio is found to be  $R^{**}=2.2\pm1.3$ assuming an age of 1 Myr, consistent with  the similar ratio predicted by the system IMF proposed by \citet{chabrierreview}. 
The ratio is  similar to the ratios observed towards the  Orion Nebula Cluster and IC 348 as well as the ratio derived in the 28 square degree survey of Taurus by \citet{guieu}. 
\end{abstract}


\section{Introduction}

The shape of the IMF and whether or not it is  universal have been central  questions in astrophysics for more than 50 years. 
The shape of the stellar part of the IMF has been examined observationally in some detail. 
 Above 0.5 M$_\odot$, the IMF derived for field stars is well characterized by a Salpeter power law, $dN/dM \propto M^{-\alpha_1}, \alpha_1=2.35$ \citep{kro02}. 
At this characteristic mass, the slope of the power law changes to $\alpha_2=1.35$ and for masses  below 0.1 M$_\odot$ the IMF flattens further \citep{kro02,chabrierreview}, although the slope is very uncertain \citep{allen}. 
To search for variations in the IMF as a function of environment, one has to look at individual star--forming episodes, such as star clusters or associations. 
Young clusters in particular are useful for these studies because the more massive stars are still present in the cluster and the clusters have not experienced significant dynamical mass segregation. 
Recently, the IMF studies have been extended deep  into the brown  dwarf regime in several star forming regions, most notably  the Orion Nebula Cluster \citep[ONC, e.g.][]{carphil,muench,slesnick}, IC 348 \citep[e.g.][]{najita00,luhman2}, and Taurus \citep[e.g.][]{briceno,luhman,guieu}.

In order to search for variations in the sub--stellar IMF as a function of environment, a larger ensemble of young clusters is needed. 
Here, we present a study of the IMF in the Mon R2 cluster. 
The embedded cluster associated with Mon R2 was originally discovered by \citet{beckwith} and is  located at a distance of 830$\pm$50 pc \citep{herbst}. 
The most massive star in the cluster is $\sim$ 10 M$_\odot$ \citep{carpenter,massi}. 
\citet{carpenter} estimated  Mon R2 to contain at  least 475 stars within a 3.6$\times$3.6 pc region, significantly less than the $\sim$ 3500 stars found within a 2.5 pc radius of the ONC \citep{hillenbrand98}. 
IC 348 contains $\sim$ 400 members  including brown dwarfs \citep{luhman2}. 
Thus, Mon R2 has a richness  intermediate between the  two well studied clusters. 
The molecular cloud has been shown by \citet{choi} to have a central density of $n_c\sim10^7\mathrm{cm}^{-3}$, much higher than typical densities in the Taurus molecular cloud cores ($n_c\sim10^5\mathrm{cm}^{-3}$, see e.g. \citet{onishi}). 
If the density of the molecular cores have an impact on the IMF as suggested by e.g. \citet{batebonnell} and \citet{goodwin}, we would expect the IMF in Mon R2 to have relatively more brown dwarfs than Taurus. 

Previously, \citet{carpenter} imaged   Mon R2  down to the brown dwarf limit and obtained spectra for a subset of the stars.  
They established the  age of the cluster to be $\le$ 3 Myr, and  that the  majority of  the low mass  stars  clustered around the  1 Myr isochrone. 
Adopting an age of 1 Myr for the stellar population, they found the ratio of high mass to low mass stars N(1--10 M$_\odot$)/N(0.1--1 M$_\odot$)$=0.1$, which is   consistent with other nearby star forming regions \citep{meyer00} and  with a \citet{miller} field star IMF. 
Despite the fact that Mon R2 is relatively nearby, the brown dwarf content  has not  been investigated in a systematic manner. 
Here, we present HST/NICMOS 2 near--infrared imaging of the central 0.24pc$\times$0.24pc (1\mbox{$^\prime$} $\times$ 1\mbox{$^\prime$}) of the cluster. 
Our main goal is to constrain the low--mass stellar and  brown dwarf content in the cluster and compare the results to those  derived for  other  young clusters and the field. 
In addition, we estimate the fraction of stars with disks within our surveyed region.

The paper is structured as follows: 
In Section 2, we present the observations, the data reduction and the photometry. 
Section 3 presents the  results from the color--magnitude and color--color diagrams for the cluster. 
We  place objects in the HR diagram based on the estimate of the effective temperature derived from the F160W and F165M band observations. 
In Section 4, we  discuss our constraints on the IMF and disk frequency based on  extinction limited samples constructed from our observations. 
Section 5 compares our derived ratios with the similar ratios for other star forming regions. 
Finally, we conclude in Section 6 with a summary of our results.


\section{Observations and reduction}
\subsection{The dataset}
HST/NICMOS2 observations of the central 1\arcmin$\times$1\arcmin\ of Mon R2 were obtained in Cycle 7 under program number 7417.  
The field of view for NICMOS 2 is 19\farcs2 square and the pixel scale is 0\farcs075. 
A $4\times4$ grid was observed in the F110W, F160W, F165M, and F207M filters, where each position in the grid was observed twice in each filter with a dither offset of $\sim$ 20 pixels in both RA and DEC. 
The total integration time per position in the grid  is 512 seconds for F110W, 352 seconds for F160W, 576 seconds for F165M, and 288 seconds for F207M. 
All the images were obtained in non--destructive readout mode with  a total of 16 readouts in each exposure. 
 The full width at half maxima of the point spread function are  0\farcs12, 0\farcs15, and  0\farcs20,  for the F110W, F160W, and F207M bands respectively. 
 Table~\ref{obs} contains the observing log.

\begin{table}
\begin{center}
\caption{Observing log for the  NICMOS2 observations.}
\label{obs}

\begin{tabular}{crrr}
\tableline\tableline
Filter & Obs. date & Exp. time  \\
\tableline
F110W & Dec 1 1997 & 2$\times$256\\
F160W & Dec 1 1997 & 2$\times$176\\
F165M & Dec 5 1997 & 2$\times$288\\
F207M & Dec 5 1997 & 2$\times$144\\
\tableline
\end{tabular}
\end{center}
\end{table}

\begin{figure*}
\includegraphics[scale=0.78,angle=90]{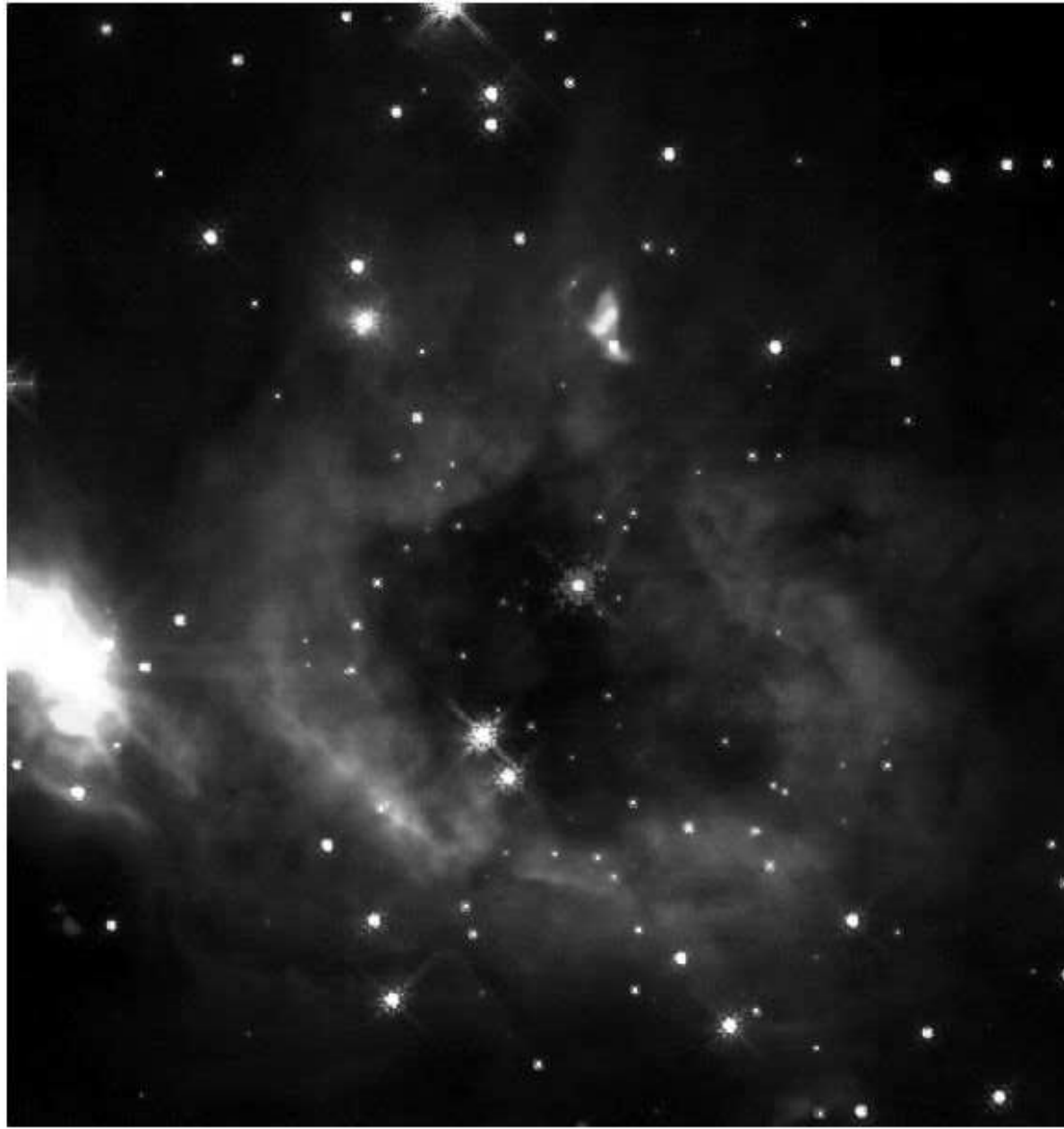}
\caption{
HST/NICMOS 2 F110W (blue), F160W (green), and F207M (red) color-composite of the central 1$\times$1\arcmin of Mon R2. 
The faintest (red) stars seen have a magnitude of $m_{F207M}\sim$19.5 mag.  North is up  and east left. The coordinates of the bright red source in the center of the image is (RA,DEC)=(06:07:45.77,-06:22:53.6)}
\label{color}
\end{figure*}

The images were processed using a combination of IRAF scripts, C programs, and IDL scripts. The basic reduction was performed using the NICRED  procedure \citep{lehar} which  performs linearity corrections, cosmic ray rejection,  determines photon rates, does dark subtraction, and flat fielding.  
Synthetic darks were used based on the software NICRED. 
The pedestal effect was removed from each frame also using NICRED. 
Bad pixels were identified and corrected using the IRAF task {\tt fixpix}. 
Finally, for each filter, the images were registered and a mosaic was created using the average of  the two dithered exposures at each position in the mosaic. 
A color composite of the observed region is shown in Fig.~\ref{color}, where F110W is  blue, F160W is green, and F207M red.  

\clearpage

\clearpage

\subsection{Source detection and photometry}
Point sources in each of the mosaics were identified using the IRAF implementation of {\tt daofind} with a 5$\sigma$ detection threshold. 
False detections in the form of noise peaks or structures in the point spread function were eliminated by comparing the source lists from each filter  and a visual inspection of each source detected by {\tt daofind}. 
Since the location of features in the PSF for a given star is dependent on wavelength, the use of different filters helps in separating faint stars next to bright stars from structure in the bright star PSF.

Aperture photometry was performed using the {\tt apphot} package in IRAF. 
An aperture of 3.5 pixel radius was used to measure the flux of the object and the background was measured in an annulus  from 10 to 15 pixels. 
For 10 objects an annulus of 2 pixel was used to measure the object flux because they were close to  bright stars. 
The choice of a relatively large aperture for the majority of the sources was to limit photometric errors due to the variations in the PSF across the field of view of the NIC2 camera. 
Some 40 stars with no (or very little) nebulosity associated with them were used to determine aperture corrections for the photometry in each filter. 
The NICMOS photometry was measured relative to a 10 pixel radius aperture and is presented in the VEGAMAG system. 
The astrometry was done  relative to \cite{carpenter}. 
Some 25 stars in common were used to fit a second order polynomial in both RA and DEC. The final RMS was 0\farcs1. 
A comparison with 2MASS showed an offset by 0\farcs5 in RA and -0\farcs3 in DEC relative to our astrometry. 
We corrected the astrometry for this offset and the RMS compared to 2MASS is then  0\farcs2 for 8 relatively bright isolated sources. 
Table~2 presents the photometry and coordinates for the 181 sources detected together with 2MASS IDs where available.  
Where more than one source was within an arcsecond in the HST/NICMOS survey, we have chosen the brightest source in our observations as the 2MASS source. 
In four cases, two sources of roughly equal brightness in the NICMOS data were identified as one object by 2MASS.  Both objects have been given the 2MASS ID and are marked with an $*$. 
\clearpage
\begin{deluxetable}{rrrrrrrrrr}
\tabletypesize{\scriptsize}

\tablehead{\colhead{ID} & \colhead{RA (2000)} & \colhead{DEC (2000)} & \colhead{F110W} & \colhead{F160W} & \colhead{F165M}  & \colhead{F207M} & \colhead{2MASS ID}}
\tablewidth{0pt}
\tablecaption{The HST/NICMOS photometry for Mon R2.} 
\label{phot}
\startdata

1 & 6:07:44.98 & -6:23:25.8 & 23.11$\pm$ 0.51 & 17.32 $\pm$ 0.05 & 17.10 $\pm$0.05 & 14.47 $\pm$ 0.06&  06074492-0623257  & \\
2 & 6:07:45.63 & -6:23:25.1 & 24.55$\pm$ 1.21 & 19.77 $\pm$ 0.07 & 19.41 $\pm$0.07 & 17.07 $\pm$ 0.07& & \\
3 & 6:07:43.89 & -6:23:26.8 & 18.58$\pm$ 0.03 & 16.20 $\pm$ 0.04 & 15.90 $\pm$0.04 & 15.48 $\pm$ 0.06& & \\
4 & 6:07:44.82 & -6:23:25.8 & 17.34$\pm$ 0.03 & 14.77 $\pm$ 0.03 & 14.55 $\pm$0.03 & 13.89 $\pm$ 0.05& 06074483-0623260  &  \\
5 & 6:07:44.26 & -6:23:25.5 & 16.12$\pm$ 0.04 & 13.81 $\pm$ 0.04 & 13.58 $\pm$0.04 & 13.04 $\pm$ 0.06& 06074426-0623256 & \\
6 & 6:07:47.67 & -6:23:22.0 & 20.17$\pm$ 0.04 & 18.45 $\pm$ 0.04 & 18.15 $\pm$0.04 & 17.90 $\pm$ 0.08& & \\
7 & 6:07:46.08 & -6:23:21.8 & 18.62$\pm$ 0.03 & 15.92 $\pm$ 0.03 & 15.68 $\pm$0.03 & 15.01 $\pm$ 0.05& & \\
8 & 6:07:44.98 & -6:23:21.9 & 21.13$\pm$ 0.05 & 17.56 $\pm$ 0.03 & 17.29 $\pm$0.03 & 16.17 $\pm$ 0.05& & \\
9 & 6:07:47.23 & -6:23:18.8 & 20.88$\pm$ 0.05 & 18.74 $\pm$ 0.04 & 18.51 $\pm$0.04 & 18.03 $\pm$ 0.08& & \\
10 & 6:07:44.53 & -6:23:21.4 & 17.75$\pm$ 0.04 & 14.96 $\pm$ 0.04 & 14.67 $\pm$0.04 & 13.63 $\pm$ 0.06& 06074454-0623214 & \\
11 & 6:07:45.31 & -6:23:20.2 & 15.51$\pm$ 0.04 & 12.99 $\pm$ 0.04 & 12.76 $\pm$0.04 & 12.13 $\pm$ 0.06& 06074531-0623202 & \\
12 & 6:07:43.82 & -6:23:21.5 & 21.76$\pm$ 0.11 & 18.46 $\pm$ 0.04 & 18.14 $\pm$0.04 & 17.08 $\pm$ 0.07& & \\
13 & 6:07:45.21 & -6:23:19.5 & 18.62$\pm$ 0.04 & 16.35 $\pm$ 0.04 & 16.06 $\pm$0.04 & 15.62 $\pm$ 0.07& & \\
14 & 6:07:46.64 & -6:23:17.3 & 15.26$\pm$ 0.04 & 12.80 $\pm$ 0.04 & 12.90 $\pm$0.04 & 12.09 $\pm$ 0.06&06074665-0623175 & \\
15 & 6:07:45.68 & -6:23:17.7 & 18.91$\pm$ 0.04 & 16.26 $\pm$ 0.04 & 16.02 $\pm$0.04 & 15.37 $\pm$ 0.07& & \\
16 & 6:07:43.96 & -6:23:18.5 & 17.00$\pm$ 0.04 & 14.52 $\pm$ 0.04 & 14.29 $\pm$0.04 & 13.71 $\pm$ 0.06&06074396-0623184 & \\
17 & 6:07:44.10 & -6:23:18.2 & 21.25$\pm$ 0.06 & 18.47 $\pm$ 0.04 & 18.28 $\pm$0.04 & 17.26 $\pm$ 0.07& & \\
18 & 6:07:45.49 & -6:23:16.0 & 16.32$\pm$ 0.04 & 14.06 $\pm$ 0.04 & 13.84 $\pm$0.04 & 13.43 $\pm$ 0.06&06074549-0623160 & \\
19 & 6:07:44.95 & -6:23:16.4 & 22.56$\pm$ 0.16 & 19.34 $\pm$ 0.05 & 19.06 $\pm$0.05 & 17.05 $\pm$ 0.05& & \\
20 & 6:07:46.31 & -6:23:13.7 & 18.43$\pm$ 0.04 & 16.00 $\pm$ 0.04 & 15.72 $\pm$0.04 & 15.21 $\pm$ 0.07&06074631-0623128 & \\
21 & 6:07:44.62 & -6:23:15.3 & 22.75$\pm$ 0.15 & 17.75 $\pm$ 0.04 & 17.42 $\pm$0.04 & 15.56 $\pm$ 0.06& & \\
22 & 6:07:45.65 & -6:23:14.1 & 18.46$\pm$ 0.04 & 16.29 $\pm$ 0.04 & 16.06 $\pm$0.04 & 15.71 $\pm$ 0.07& & \\
23 & 6:07:44.37 & -6:23:15.4 & ....  & 20.90 $\pm$ 0.17 & 20.79 $\pm$0.18 & 18.54 $\pm$ 0.12& & \\
24 & 6:07:44.48 & -6:23:15.2 & ....  & 20.43 $\pm$ 0.12 & 20.40 $\pm$0.14 & 17.49 $\pm$ 0.08& & \\
25 & 6:07:47.74 & -6:23:11.8 & 16.98$\pm$ 0.04 & 15.01 $\pm$ 0.04 & 14.77 $\pm$0.04 & 14.45 $\pm$ 0.06& & \\
26 & 6:07:47.12 & -6:23:12.2 & 21.33$\pm$ 0.07 & 18.75 $\pm$ 0.05 & 18.47 $\pm$0.05 & 17.74 $\pm$ 0.09& & \\
27 & 6:07:46.69 & -6:23:12.6 & 15.84$\pm$ 0.04 & 13.58 $\pm$ 0.04 & 13.35 $\pm$0.04 & 12.87 $\pm$ 0.06&06074669-0623125 & \\
28 & 6:07:44.80 & -6:23:14.4 & 16.95$\pm$ 0.03 & 13.79 $\pm$ 0.03 & 13.59 $\pm$0.03 & 12.59 $\pm$ 0.05&06074482-0623144 & \\
29 & 6:07:46.33 & -6:23:12.1 & 20.55$\pm$ 0.05 & 16.52 $\pm$ 0.04 & 16.21 $\pm$0.04 & 14.37 $\pm$ 0.06& & \\
30 & 6:07:45.46 & -6:23:12.9 & 21.62$\pm$ 0.14 & 18.59 $\pm$ 0.08 & 18.33 $\pm$0.08 & 17.50 $\pm$ 0.12& & \\
31 & 6:07:47.26 & -6:23:10.5 & 24.16$\pm$ 0.67 & 21.84 $\pm$ 0.69 & 22.11 $\pm$1.15 & 18.99 $\pm$ 0.24& & \\
32 & 6:07:43.95 & -6:23:13.0 & 21.38$\pm$ 0.07 & 16.62 $\pm$ 0.04 & 16.40 $\pm$0.04 & 14.10 $\pm$ 0.06&06074393-0623130 & \\
33 & 6:07:46.23 & -6:23:10.4 & ....  &  ....   &  .... & 17.91 $\pm$ 0.18& & \\
34 & 6:07:45.73 & -6:23:10.9 & 20.83$\pm$ 0.05 & 17.38 $\pm$ 0.03 & 17.06 $\pm$0.03 & 15.79 $\pm$ 0.05&06074579-0623113 & \\
35 & 6:07:45.11 & -6:23:10.8 & 20.79$\pm$ 0.06 & 16.52 $\pm$ 0.04 & 16.07 $\pm$0.04 & 14.16 $\pm$ 0.06&06074593-0623110 & \\
36 & 6:07:45.80 & -6:23:09.6 & 20.79$\pm$ 0.04 & 16.93 $\pm$ 0.02 & 16.63 $\pm$0.02 & 15.24 $\pm$ 0.04&06074500-0623113 & \\
37 & 6:07:45.96 & -6:23:09.2 & 21.37$\pm$ 0.14 & 17.80 $\pm$ 0.04 & 17.63 $\pm$0.04 & 15.67 $\pm$ 0.04& & \\
38 & 6:07:46.16 & -6:23:09.0 & ....  & 19.18 $\pm$ 0.09 & 19.02 $\pm$0.09 & 16.35 $\pm$ 0.06& & \\
39 & 6:07:46.86 & -6:23:07.9 & 16.73$\pm$ 0.02 & 14.64 $\pm$ 0.02 & 14.39 $\pm$0.02 & 13.90 $\pm$ 0.03&06074686-0623076 & \\
40 & 6:07:43.99 & -6:23:10.6 & 19.51$\pm$ 0.03 & 16.63 $\pm$ 0.03 & 16.36 $\pm$0.03 & 15.60 $\pm$ 0.05&   & \\
41 & 6:07:43.68 & -6:23:10.4 & 17.50$\pm$ 0.03 & 15.12 $\pm$ 0.03 & 14.84 $\pm$0.03 & 14.29 $\pm$ 0.05&06074370-0623107 & \\
42 & 6:07:45.16 & -6:23:08.7 & 20.17$\pm$ 0.04 & 16.42 $\pm$ 0.03 & 16.16 $\pm$0.03 & 14.96 $\pm$ 0.05&06074516-0623103 & \\
43 & 6:07:45.42 & -6:23:08.2 & 19.23$\pm$ 0.03 & 15.62 $\pm$ 0.03 & 15.11 $\pm$0.03 & 13.86 $\pm$ 0.05&06074542-0623092 & \\
44 & 6:07:44.81 & -6:23:08.4 & 24.04$\pm$ 0.68 & 19.33 $\pm$ 0.04 & 19.08 $\pm$0.03 & 16.69 $\pm$ 0.04& & \\
45 & 6:07:45.89 & -6:23:07.3 & 22.60$\pm$ 0.16 & 19.47 $\pm$ 0.04 & 19.21 $\pm$0.04 & 17.65 $\pm$ 0.05& & \\
46 & 6:07:46.64 & -6:23:05.9 & 20.61$\pm$ 0.12 & 17.02 $\pm$ 0.07 & 16.78 $\pm$0.07 & 15.41 $\pm$ 0.08&06074661-0623058 & \\
47 & 6:07:47.25 & -6:23:05.1 & 21.95$\pm$ 0.15 & 19.59 $\pm$ 0.08 & 19.27 $\pm$0.08 & 18.18 $\pm$ 0.15& & \\
48 & 6:07:46.62 & -6:23:05.7 & 21.01$\pm$ 0.19 & 18.23 $\pm$ 0.09 & 17.94 $\pm$0.09 & 16.45 $\pm$ 0.08& & \\
49 & 6:07:45.64 & -6:23:06.5 & 20.92$\pm$ 0.05 & 16.55 $\pm$ 0.03 & 16.23 $\pm$0.03 & 14.78 $\pm$ 0.05& & \\
50 & 6:07:47.84 & -6:23:03.7 & 16.52$\pm$ 0.03 & 13.82 $\pm$ 0.03 & 13.58 $\pm$0.03 & 12.82 $\pm$ 0.05&06074784-0623038 & \\
51 & 6:07:45.03 & -6:23:06.5 & 20.49$\pm$ 0.04 & 16.78 $\pm$ 0.03 & 16.68 $\pm$0.03 & 15.49 $\pm$ 0.05&06074503-0623060 & \\
52 & 6:07:45.08 & -6:23:06.0 & 21.60$\pm$ 0.11 & 17.03 $\pm$ 0.04 & 16.71 $\pm$0.04 & 15.18 $\pm$ 0.07& & \\
53 & 6:07:46.13 & -6:23:04.5 & 17.19$\pm$ 0.04 & 12.83 $\pm$ 0.04 & 12.57 $\pm$0.04 & 10.78 $\pm$ 0.06&06074615-0623046 & \\
54 & 6:07:44.62 & -6:23:05.3 & 19.96$\pm$ 0.04 & 16.66 $\pm$ 0.04 & 16.39 $\pm$0.04 & 15.25 $\pm$ 0.07&06074470-0623044 & \\
55 & 6:07:48.07 & -6:23:01.9 & 23.92$\pm$ 0.81 & 18.65 $\pm$ 0.04 & 18.34 $\pm$0.04 & 16.13 $\pm$ 0.05&06074807-0623017 & \\
56 & 6:07:44.80 & -6:23:05.1 & 17.33$\pm$ 0.04 & 15.51 $\pm$ 0.04 & 15.27 $\pm$0.04 & 15.19 $\pm$ 0.07&06074470-0623044 & \\
57 & 6:07:45.88 & -6:23:03.5 & 22.96$\pm$ 0.38 & 18.92 $\pm$ 0.04 & 18.56 $\pm$0.04 & 16.06 $\pm$ 0.05& & \\
58 & 6:07:47.67 & -6:23:01.1 & 18.91$\pm$ 0.04 & 16.95 $\pm$ 0.07 & 16.64 $\pm$0.07 & 16.56 $\pm$ 0.28& & \\
59 & 6:07:45.26 & -6:23:03.2 & 22.02$\pm$ 0.10 & 17.65 $\pm$ 0.04 & 17.58 $\pm$0.04 & 15.31 $\pm$ 0.06& & \\
60 & 6:07:44.62 & -6:23:03.9 & ....  & 19.66 $\pm$ 0.14 & 19.21 $\pm$0.11 & 16.52 $\pm$ 0.08& & \\
61 & 6:07:46.22 & -6:23:01.9 & 12.34$\pm$ 0.04 & 11.57 $\pm$ 0.04 & 11.45 $\pm$0.04 & 11.63 $\pm$ 0.06& 06074622-0623022 & \\
62 & 6:07:46.02 & -6:23:01.6 & ....  & 18.52 $\pm$ 0.07 & 18.31 $\pm$0.06 & 14.78 $\pm$ 0.06& & \\
63 & 6:07:45.67 & -6:23:01.9 & 23.57$\pm$ 0.47 & 19.77 $\pm$ 0.09 & 19.48 $\pm$0.09 & 17.01 $\pm$ 0.08& & \\
64 & 6:07:45.85 & -6:23:01.4 & 25.11$\pm$ 3.73 & 21.61 $\pm$ 0.29 & 20.85 $\pm$0.18 & 18.57 $\pm$ 0.13& & \\
65 & 6:07:45.99 & -6:23:01.1 & 21.89$\pm$ 0.17 & 20.78 $\pm$ 0.19 & 20.26 $\pm$0.17 & 18.08 $\pm$ 0.11& & \\
66 & 6:07:45.04 & -6:23:01.8 & 21.59$\pm$ 0.18 & 18.76 $\pm$ 0.11 & 18.43 $\pm$0.09 & 17.49 $\pm$ 0.13&06074495-0623010 & \\
67 & 6:07:46.16 & -6:23:00.4 & 22.44$\pm$ 0.38 & 18.79 $\pm$ 0.07 & 18.41 $\pm$0.06 & 16.12 $\pm$ 0.07& & \\
68 & 6:07:45.65 & -6:23:00.8 & 26.05$\pm$ 3.84 & 21.88 $\pm$ 0.56 & 21.53 $\pm$0.46 & 18.51 $\pm$ 0.17& & \\
69 & 6:07:46.01 & -6:23:00.3 & 23.67$\pm$ 1.21 &  ....  & 23.56 $\pm$4.82 & 21.81 $\pm$ 3.48& & \\
70 & 6:07:45.98 & -6:22:59.9 & 23.32$\pm$ 0.82 &  ....  & .... & 19.46 $\pm$ 0.46& & \\
71 & 6:07:45.71 & -6:23:00.0 & 22.59$\pm$ 0.22 & 18.03 $\pm$ 0.03 & 17.69 $\pm$0.03 & 15.05 $\pm$ 0.05& & \\
72 & 6:07:45.64 & -6:22:59.4 & 23.07$\pm$ 0.23 & 21.15 $\pm$ 0.29 & 20.38 $\pm$0.16 & 19.48 $\pm$ 0.43& & \\
73 & 6:07:46.75 & -6:22:57.6 & 21.64$\pm$ 0.17 & 18.45 $\pm$ 0.10 & 18.07 $\pm$0.10 & 16.49 $\pm$ 0.10& & \\
74 & 6:07:46.72 & -6:22:57.5 & 20.21$\pm$ 0.08 & 16.54 $\pm$ 0.07 & 16.19 $\pm$0.07 & 14.95 $\pm$ 0.07&06074674-0622562 & \\
75 & 6:07:47.53 & -6:22:56.5 & 17.33$\pm$ 0.04 & 15.11 $\pm$ 0.04 & 14.84 $\pm$0.04 & 14.35 $\pm$ 0.07& & \\
76 & 6:07:46.89 & -6:22:57.0 & 20.58$\pm$ 0.06 & 18.54 $\pm$ 0.04 & 18.25 $\pm$0.04 & 17.72 $\pm$ 0.08& & \\
77 & 6:07:47.84 & -6:22:56.1 & 14.23$\pm$ 0.04 & 10.43 $\pm$ 0.04 & 10.16 $\pm$0.04 & 8.21 $\pm$ 0.06&06074786-0622559 & \\
78 & 6:07:47.74 & -6:22:56.1 & 17.89$\pm$ 0.05 & 14.75 $\pm$ 0.04 & 14.45 $\pm$0.04 & 13.04 $\pm$ 0.06& & \\
79 & 6:07:46.28 & -6:22:57.1 & ....  & 18.05 $\pm$ 0.04 & 18.22 $\pm$0.05 & 15.31 $\pm$ 0.06& & \\
80 & 6:07:47.89 & -6:22:55.2 & 15.25$\pm$ 0.04 & 11.91 $\pm$ 0.04 & 11.68 $\pm$0.04 & 9.73 $\pm$ 0.06& & \\
81 & 6:07:46.43 & -6:22:56.3 & 22.72$\pm$ 0.25 & 19.58 $\pm$ 0.07 & 19.19 $\pm$0.06 & 17.69 $\pm$ 0.08& & \\
82 & 6:07:47.68 & -6:22:55.0 & 15.70$\pm$ 0.04 & 13.66 $\pm$ 0.04 & 13.34 $\pm$0.04 & 12.78 $\pm$ 0.06& & \\
83 & 6:07:46.89 & -6:22:55.6 & 19.27$\pm$ 0.04 & 18.27 $\pm$ 0.04 & 18.07 $\pm$0.04 & 18.15 $\pm$ 0.08& & \\
84 & 6:07:45.02 & -6:22:56.9 & 22.39$\pm$ 0.12 & 18.05 $\pm$ 0.05 & 17.76 $\pm$0.05 & 16.14 $\pm$ 0.07&06074488-0622563 & \\
85 & 6:07:47.89 & -6:22:54.0 & 17.08$\pm$ 0.04 & 13.85 $\pm$ 0.04 & 13.71 $\pm$0.04 & 12.42 $\pm$ 0.08& & \\
86 & 6:07:46.69 & -6:22:54.8 & 19.66$\pm$ 0.05 & 15.88 $\pm$ 0.04 & 15.61 $\pm$0.04 & 14.42 $\pm$ 0.06& & \\
87 & 6:07:47.38 & -6:22:53.9 & 17.61$\pm$ 0.04 & 15.05 $\pm$ 0.04 & 14.81 $\pm$0.04 & 14.03 $\pm$ 0.06& & \\
88 & 6:07:45.45 & -6:22:55.7 & 23.27$\pm$ 0.34 & 19.46 $\pm$ 0.09 & 19.08 $\pm$0.08 & 17.65 $\pm$ 0.11& & \\
89 & 6:07:45.72 & -6:22:55.0 & 23.12$\pm$ 0.30 & 18.90 $\pm$ 0.04 & 18.62 $\pm$0.04 & 15.44 $\pm$ 0.05& & \\
90 & 6:07:47.67 & -6:22:53.0 & 20.85$\pm$ 0.08 & 18.02 $\pm$ 0.10 & 17.74 $\pm$0.10 & 16.04 $\pm$ 0.09& & \\
91 & 6:07:46.03 & -6:22:54.5 & 25.26$\pm$ 2.81 & 19.39 $\pm$ 0.09 & 19.00 $\pm$0.08 & 16.60 $\pm$ 0.07& & \\
92 & 6:07:46.37 & -6:22:54.0 & 22.33$\pm$ 0.20 & 19.59 $\pm$ 0.09 & 19.24 $\pm$0.08 & 17.83 $\pm$ 0.09& & \\
93 & 6:07:46.10 & -6:22:54.1 & ....  & 18.57 $\pm$ 0.04 & 18.40 $\pm$0.04 & 15.39 $\pm$ 0.05& & \\
94 & 6:07:45.64 & -6:22:54.3 & 24.30$\pm$ 0.67 & 18.46 $\pm$ 0.04 & 18.30 $\pm$0.04 & 15.97 $\pm$ 0.05& & \\
95 & 6:07:44.14 & -6:22:55.5 & 23.41$\pm$ 0.29 & 20.73 $\pm$ 0.17 & 20.32 $\pm$0.15 & 19.14 $\pm$ 0.24& & \\
96 & 6:07:45.80 & -6:22:53.3 & 19.54$\pm$ 0.02 & 14.59 $\pm$ 0.02 & 14.52 $\pm$0.02 & 9.77 $\pm$ 0.03& & \\
97 & 6:07:46.01 & -6:22:53.0 & 23.28$\pm$ 0.36 & 19.93 $\pm$ 0.09 & 19.75 $\pm$0.08 & 16.99 $\pm$ 0.06& & \\
98 & 6:07:46.60 & -6:22:52.4 & 19.67$\pm$ 0.03 & 16.12 $\pm$ 0.03 & 15.77 $\pm$0.03 & 14.69 $\pm$ 0.05& & \\
99 & 6:07:45.69 & -6:22:52.4 & 22.92$\pm$ 0.31 & 18.80 $\pm$ 0.04 & 18.87 $\pm$0.04 & 15.42 $\pm$ 0.04& & \\
100 & 6:07:45.75 & -6:22:51.7 & 22.53$\pm$ 0.19 & 20.97 $\pm$ 0.19 & 21.18 $\pm$0.34 & 16.68 $\pm$ 0.07& & \\
101 & 6:07:46.42 & -6:22:50.8 & 22.43$\pm$ 0.14 & 19.91 $\pm$ 0.13 & 19.75 $\pm$0.13 & 18.33 $\pm$ 0.19& & \\
102 & 6:07:44.54 & -6:22:52.6 & 23.87$\pm$ 0.60 & 20.54 $\pm$ 0.13 & 20.20 $\pm$0.12 & 17.76 $\pm$ 0.06&06074443-0622523 & \\
103 & 6:07:46.47 & -6:22:50.4 & 21.84$\pm$ 0.08 & 17.90 $\pm$ 0.03 & 17.60 $\pm$0.03 & 16.03 $\pm$ 0.05& & \\
104 & 6:07:43.64 & -6:22:52.5 & 16.40$\pm$ 0.04 & 13.84 $\pm$ 0.04 & 13.64 $\pm$0.04 & 12.80 $\pm$ 0.06&06074361-0622519* & \\
105 & 6:07:46.38 & -6:22:49.5 & 21.53$\pm$ 0.08 & 18.22 $\pm$ 0.04 & 17.83 $\pm$0.04 & 16.50 $\pm$ 0.06& & \\
106 & 6:07:46.18 & -6:22:49.6 & 23.01$\pm$ 0.24 & 19.87 $\pm$ 0.16 & 19.50 $\pm$0.14 & 17.52 $\pm$ 0.16& & \\
107 & 6:07:45.60 & -6:22:50.1 & 21.99$\pm$ 0.10 & 17.74 $\pm$ 0.03 & 17.48 $\pm$0.03 & 15.70 $\pm$ 0.05& & \\
108 & 6:07:46.26 & -6:22:49.4 & 23.11$\pm$ 0.28 & 17.94 $\pm$ 0.05 & 17.61 $\pm$0.05 & 15.46 $\pm$ 0.07& & \\
109 & 6:07:45.70 & -6:22:49.4 & 21.68$\pm$ 0.09 & 17.13 $\pm$ 0.03 & 16.79 $\pm$0.03 & 15.11 $\pm$ 0.05& & \\
110 & 6:07:44.10 & -6:22:50.9 & 22.68$\pm$ 0.25 & 19.63 $\pm$ 0.07 & 19.25 $\pm$0.06 & 18.04 $\pm$ 0.09& & \\
111 & 6:07:43.60 & -6:22:51.3 & 15.61$\pm$ 0.04 & 13.15 $\pm$ 0.04 & 12.93 $\pm$0.04 & 12.43 $\pm$ 0.06&06074361-0622519* & \\
112 & 6:07:45.56 & -6:22:49.3 & 22.61$\pm$ 0.19 & 17.51 $\pm$ 0.04 & 17.19 $\pm$0.04 & 14.94 $\pm$ 0.07& & \\
113 & 6:07:46.33 & -6:22:46.8 & 20.40$\pm$ 0.05 & 17.12 $\pm$ 0.04 & 16.78 $\pm$0.04 & 15.87 $\pm$ 0.07&06074636-0622467 & \\
114 & 6:07:45.30 & -6:22:47.1 & 22.72$\pm$ 0.20 & 18.70 $\pm$ 0.06 & 18.41 $\pm$0.05 & 16.19 $\pm$ 0.07& & \\
115 & 6:07:46.27 & -6:22:45.6 & 21.50$\pm$ 0.11 & 17.99 $\pm$ 0.06 & 17.72 $\pm$0.05 & 16.40 $\pm$ 0.08& & \\
116 & 6:07:46.49 & -6:22:45.0 & 22.63$\pm$ 0.17 & 17.62 $\pm$ 0.04 & 17.41 $\pm$0.04 & 15.09 $\pm$ 0.07& & \\
117 & 6:07:45.07 & -6:22:46.3 & 20.86$\pm$ 0.06 & 16.71 $\pm$ 0.04 & 16.48 $\pm$0.04 & 14.87 $\pm$ 0.06& & \\
118 & 6:07:44.97 & -6:22:46.4 & 21.44$\pm$ 0.08 & 17.52 $\pm$ 0.04 & 17.23 $\pm$0.04 & 15.92 $\pm$ 0.07& & \\
119 & 6:07:45.16 & -6:22:45.7 & 22.76$\pm$ 0.23 & 20.80 $\pm$ 0.16 & 20.37 $\pm$0.13 & 19.45 $\pm$ 0.34& & \\
120 & 6:07:43.74 & -6:22:46.7 & 16.92$\pm$ 0.09 & 14.71 $\pm$ 0.09 & 14.40 $\pm$0.09 & 14.00 $\pm$ 0.10&06074374-0622467 & \\
121 & 6:07:43.75 & -6:22:46.4 & 17.77$\pm$ 0.09 & 15.41 $\pm$ 0.09 & 15.15 $\pm$0.09 & 14.70 $\pm$ 0.10& & \\
122 & 6:07:44.45 & -6:22:44.8 & 22.07$\pm$ 0.13 & 17.42 $\pm$ 0.04 & 17.14 $\pm$0.04 & 15.49 $\pm$ 0.06& & \\
123 & 6:07:46.40 & -6:22:42.7 & 17.74$\pm$ 0.04 & 15.21 $\pm$ 0.04 & 15.00 $\pm$0.04 & 14.42 $\pm$ 0.06&06074640-0622432 & \\
124 & 6:07:43.55 & -6:22:44.6 & 19.28$\pm$ 0.09 & 16.04 $\pm$ 0.09 & 15.74 $\pm$0.09 & 14.58 $\pm$ 0.10&06074356-0622448* & \\
125 & 6:07:43.56 & -6:22:44.6 & 18.74$\pm$ 0.09 & 15.46 $\pm$ 0.09 & 15.17 $\pm$0.09 & 14.20 $\pm$ 0.10&06074356-0622448* & \\
126 & 6:07:43.90 & -6:22:44.2 & 22.41$\pm$ 0.20 & 20.15 $\pm$ 0.07 & 19.84 $\pm$0.07 & 18.93 $\pm$ 0.12& & \\
127 & 6:07:46.94 & -6:22:40.9 & 21.87$\pm$ 0.09 & 17.79 $\pm$ 0.03 & 17.49 $\pm$0.03 & 16.10 $\pm$ 0.05& & \\
128 & 6:07:46.27 & -6:22:41.0 & 23.00$\pm$ 0.27 & 20.57 $\pm$ 0.20 & 20.00 $\pm$0.16 & 18.91 $\pm$ 0.33& & \\
129 & 6:07:44.99 & -6:22:42.2 & 24.44$\pm$ 1.09 & 19.35 $\pm$ 0.07 & 18.97 $\pm$0.06 & 17.48 $\pm$ 0.08& & \\
130 & 6:07:45.81 & -6:22:41.4 & 24.99$\pm$ 1.97 & 19.06 $\pm$ 0.05 & 18.88 $\pm$0.05 & 16.53 $\pm$ 0.05& & \\
131 & 6:07:44.76 & -6:22:41.5 & 24.28$\pm$ 0.96 & 20.88 $\pm$ 0.12 & 20.82 $\pm$0.13 & 18.33 $\pm$ 0.08& & \\
132 & 6:07:47.43 & -6:22:38.5 & 24.01$\pm$ 0.44 & 23.40 $\pm$ 1.12 & 24.70 $\pm$4.58 & 18.28 $\pm$ 0.09& & \\
133 & 6:07:44.48 & -6:22:41.2 & 18.30$\pm$ 0.04 & 15.11 $\pm$ 0.04 & 14.88 $\pm$0.04 & 13.93 $\pm$ 0.06&06074449-0622414 & \\
134 & 6:07:46.36 & -6:22:38.8 & 21.17$\pm$ 0.07 & 17.37 $\pm$ 0.04 & 17.03 $\pm$0.04 & 15.79 $\pm$ 0.07& & \\
135 & 6:07:44.96 & -6:22:40.0 & 16.57$\pm$ 0.03 & 13.86 $\pm$ 0.03 & 13.90 $\pm$0.03 & 13.02 $\pm$ 0.05&06074496-0622401 & \\
136 & 6:07:46.32 & -6:22:38.4 & 18.28$\pm$ 0.02 & 15.13 $\pm$ 0.02 & 14.82 $\pm$0.02 & 13.43 $\pm$ 0.03& & \\
137 & 6:07:46.59 & -6:22:37.2 & 22.06$\pm$ 0.12 & 19.68 $\pm$ 0.10 & 19.47 $\pm$0.11 & 18.05 $\pm$ 0.12& & \\
138 & 6:07:43.70 & -6:22:40.1 & 20.22$\pm$ 0.04 & 15.90 $\pm$ 0.03 & 15.65 $\pm$0.03 & 13.46 $\pm$ 0.05& 06074371-0622402 & \\
139 & 6:07:45.78 & -6:22:37.6 & 17.95$\pm$ 0.03 & 15.52 $\pm$ 0.03 & 15.06 $\pm$0.03 & 14.51 $\pm$ 0.05& & \\
140 & 6:07:43.65 & -6:22:37.7 & 22.41$\pm$ 0.11 & 19.36 $\pm$ 0.05 & 19.13 $\pm$0.05 & 17.40 $\pm$ 0.07& & \\
141 & 6:07:46.58 & -6:22:36.8 & 22.16$\pm$ 0.14 & 20.05 $\pm$ 0.08 & 19.76 $\pm$0.08 & 18.40 $\pm$ 0.08&06074658-0622370 & \\
142 & 6:07:45.60 & -6:22:37.4 & 17.58$\pm$ 0.03 & 13.26 $\pm$ 0.03 & 13.04 $\pm$0.03 & 11.03 $\pm$ 0.05&06074563-0622380 & \\
143 & 6:07:47.01 & -6:22:35.4 & 19.83$\pm$ 0.03 & 16.74 $\pm$ 0.03 & 16.42 $\pm$0.03 & 15.62 $\pm$ 0.05& & \\
144 & 6:07:43.84 & -6:22:38.4 & 22.91$\pm$ 0.22 & 19.27 $\pm$ 0.03 & 18.79 $\pm$0.03 & 17.01 $\pm$ 0.05& & \\
145 & 6:07:47.48 & -6:22:34.4 & 22.98$\pm$ 0.17 & 20.56 $\pm$ 0.08 & 20.13 $\pm$0.06 & 19.66 $\pm$ 0.25& & \\
146 & 6:07:45.74 & -6:22:35.3 & 22.73$\pm$ 0.18 & 18.24 $\pm$ 0.02 & 18.03 $\pm$0.02 & 15.56 $\pm$ 0.03& 06074565-0622355 & \\
147 & 6:07:43.55 & -6:22:37.0 & 18.23$\pm$ 0.04 & 15.47 $\pm$ 0.04 & 15.11 $\pm$0.04 & 14.51 $\pm$ 0.06& 06074355-0622371 & \\
148 & 6:07:46.60 & -6:22:33.5 & 16.28$\pm$ 0.04 & 13.56 $\pm$ 0.04 & 13.41 $\pm$0.04 & 12.45 $\pm$ 0.06&06074660-0622335 & \\
149 & 6:07:46.60 & -6:22:33.0 & ....  & 15.54 $\pm$ 0.04 & 15.92 $\pm$0.04 & 14.35 $\pm$ 0.06& & \\
150 & 6:07:47.33 & -6:22:32.0 & ....  & 22.20 $\pm$ 0.44 & 21.99 $\pm$0.43 & 21.74 $\pm$ 1.74& & \\
151 & 6:07:45.34 & -6:22:33.8 & 20.43$\pm$ 0.05 & 17.22 $\pm$ 0.04 & 16.85 $\pm$0.04 & 15.55 $\pm$ 0.07&06074540-0622334 & \\
152 & 6:07:45.44 & -6:22:33.5 & 22.04$\pm$ 0.10 & 17.31 $\pm$ 0.04 & 17.31 $\pm$0.04 & 14.63 $\pm$ 0.06& & \\
153 & 6:07:44.11 & -6:22:34.5 & 24.27$\pm$ 0.75 & 21.32 $\pm$ 0.17 & 21.16 $\pm$0.19 & 18.91 $\pm$ 0.11& & \\
154 & 6:07:45.94 & -6:22:32.4 & 18.86$\pm$ 0.04 & 15.33 $\pm$ 0.04 & 15.03 $\pm$0.04 & 13.77 $\pm$ 0.06& & \\
155 & 6:07:47.17 & -6:22:31.2 & 16.25$\pm$ 0.04 & 13.70 $\pm$ 0.04 & 13.50 $\pm$0.04 & 12.62 $\pm$ 0.06&06074717-0622313 & \\
156 & 6:07:44.52 & -6:22:33.6 & ....  & 24.08 $\pm$ 2.36 & 25.67 $\pm$12.71 & 19.13 $\pm$ 0.16& & \\
157 & 6:07:46.13 & -6:22:29.2 & 24.96$\pm$ 1.64 & 22.35 $\pm$ 0.59 & 21.56 $\pm$0.31 & 20.36 $\pm$ 0.41& & \\
158 & 6:07:44.24 & -6:22:30.5 & 16.57$\pm$ 0.09 & 14.64 $\pm$ 0.09 & 14.45 $\pm$0.09 & 14.19 $\pm$ 0.10&06074426-0622305* & \\
159 & 6:07:44.26 & -6:22:30.3 & 15.91$\pm$ 0.09 & 14.05 $\pm$ 0.09 & 13.83 $\pm$0.09 & 13.63 $\pm$ 0.10&06074426-0622305* & \\
160 & 6:07:47.35 & -6:22:27.2 & 18.33$\pm$ 0.04 & 16.67 $\pm$ 0.04 & 16.43 $\pm$0.04 & 16.25 $\pm$ 0.07& & \\
161 & 6:07:43.99 & -6:22:30.0 & 17.17$\pm$ 0.04 & 14.53 $\pm$ 0.04 & 14.25 $\pm$0.04 & 13.62 $\pm$ 0.06&06074398-0622301 & \\
162 & 6:07:43.82 & -6:22:30.0 & 18.34$\pm$ 0.04 & 15.83 $\pm$ 0.04 & 15.54 $\pm$0.04 & 14.89 $\pm$ 0.06& & \\
163 & 6:07:44.82 & -6:22:28.9 & 21.04$\pm$ 0.05 & 17.47 $\pm$ 0.03 & 17.31 $\pm$0.03 & 16.07 $\pm$ 0.05& 06074485-0622300 & \\
164 & 6:07:45.33 & -6:22:28.0 & 15.81$\pm$ 0.04 & 14.02 $\pm$ 0.04 & 13.83 $\pm$0.04 & 13.63 $\pm$ 0.06&06074533-0622282 & \\
165 & 6:07:46.03 & -6:22:25.5 & 16.44$\pm$ 0.04 & 13.94 $\pm$ 0.04 & 13.62 $\pm$0.04 & 12.93 $\pm$ 0.06&06074602-0622242* & \\
166 & 6:07:46.40 & -6:22:24.4 & 17.79$\pm$ 0.04 & 14.75 $\pm$ 0.04 & 14.51 $\pm$0.04 & 13.60 $\pm$ 0.06&06074641-0622245 & \\
167 & 6:07:45.65 & -6:22:25.1 & 23.09$\pm$ 0.29 & 21.22 $\pm$ 0.30 & 21.07 $\pm$0.23 & 18.81 $\pm$ 0.17& & \\
168 & 6:07:47.10 & -6:22:23.4 & 25.87$\pm$ 3.04 &  ....  & 22.14 $\pm$0.43 & 19.19 $\pm$ 0.15& & \\
169 & 6:07:46.02 & -6:22:23.7 & 15.82$\pm$ 0.04 & 13.15 $\pm$ 0.04 & 12.84 $\pm$0.04 & 12.07 $\pm$ 0.06&06074602-0622242* & \\
170 & 6:07:46.29 & -6:22:23.2 & 21.42$\pm$ 0.07 & 17.86 $\pm$ 0.04 & 17.55 $\pm$0.04 & 16.39 $\pm$ 0.07& & \\
171 & 6:07:45.70 & -6:22:23.3 & 17.47$\pm$ 0.03 & 15.50 $\pm$ 0.03 & 15.23 $\pm$0.03 & 14.96 $\pm$ 0.05&06074570-0622235 & \\
172 & 6:07:47.01 & -6:22:20.7 & 17.81$\pm$ 0.04 & 15.18 $\pm$ 0.04 & 14.94 $\pm$0.04 & 14.33 $\pm$ 0.06&06074702-0622208 & \\
173 & 6:07:46.31 & -6:22:20.5 & 21.90$\pm$ 0.15 & 19.27 $\pm$ 0.09 & 18.98 $\pm$0.08 & 17.39 $\pm$ 0.10& & \\
174 & 6:07:45.77 & -6:22:20.5 & 18.42$\pm$ 0.03 & 15.79 $\pm$ 0.03 & 15.50 $\pm$0.03 & 14.79 $\pm$ 0.05& & \\
175 & 6:07:44.14 & -6:22:22.1 & ....  & 21.88 $\pm$ 0.28 & 20.65 $\pm$0.12 & 19.21 $\pm$ 0.14& & \\
176 & 6:07:45.18 & -6:22:20.9 & 21.93$\pm$ 0.12 & 19.29 $\pm$ 0.052 & 19.18 $\pm$0.05 & 18.92 $\pm$ 0.14& & \\
177 & 6:07:46.62 & -6:22:19.4 & 23.10$\pm$ 0.33 & 20.45 $\pm$ 0.18 & 20.31 $\pm$0.19 & 18.09 $\pm$ 0.10& & \\
178 & 6:07:47.52 & -6:22:18.3 & 19.78$\pm$ 0.04 & 16.11 $\pm$ 0.04 & 15.79 $\pm$0.04 & 14.04 $\pm$ 0.06& 06074753-0622184 & \\
179 & 6:07:44.76 & -6:22:20.8 & 22.27$\pm$ 0.12 & 17.41 $\pm$ 0.03 & 17.12 $\pm$0.03 & 15.39 $\pm$ 0.05& & \\
180 & 6:07:46.41 & -6:22:18.8 & 23.79$\pm$ 0.76 & 18.48 $\pm$ 0.05 & 18.27 $\pm$0.05 & 16.28 $\pm$ 0.07& & \\
181 & 6:07:46.57 & -6:22:18.5 & 17.77$\pm$ 0.04 & 15.12 $\pm$ 0.04 & 14.83 $\pm$0.04 & 14.14 $\pm$ 0.06&06074656-0622188 & \\

\enddata
\end{deluxetable}
\clearpage

The photometry has been converted into the CIT system using 41 stars in common with \citet{carpenter}. 
The stars covered a color range $0.4 \le m_{F160W}-m_{F207M} \le 3.2 $, where $m_{F160W}$ and $m_{F207M}$ are the magnitudes in the F160W and F207M bands, respectively. 
Least--square fits were performed to convert the NICMOS photometry into the CIT system, resulting in fits  given by: 
\begin{eqnarray}
m_J & = & m_{F110W}-(0.789\pm 0.025) \nonumber\\ & - & (0.200\pm 0.027)\times (m_{F160W}-m_{F207M})\\
m_H & = & m_{F160W}-(0.089\pm0.015) \nonumber\\ & - &(0.264\pm0.013)\times(m_{F160W}-m_{F207M})\\
m_K & =& m_{F207M}-(0.214\pm0.018)\nonumber\\ & - & (0.306\pm0.012)\times(m_{F160M}-m_{F207M}). 
\end{eqnarray}
The photometry reaches roughly 2.5 mag deeper in the J band than the observations by \citet{carpenter}. 
 
\subsection{Completeness limits} 
Artificial star experiments have been performed in order to assess the completeness limits of the data.
Synthetic PSFs have been created for each filter, using the Tinytim software \citep{krist}, and a total of 160 artificial stars have been placed in each mosaic. 
The fraction of artificial stars recovered as a function of magnitude in each filter is presented in Table~\ref{completeness}. 
All the recovery fractions are for a 5$\sigma$ threshold and were subject to visual inspection in a way identical to that used to detect sources in our survey. 
We have adopted the 90\%\ completeness limit in this paper, which  corresponds to the following limiting magnitudes: $m_{F110W}$=21.5 mag, $m_{F160W}$=20.5 mag, and m$_{F207M}$=18.0 mag. 
\begin{table}
\begin{center}
\caption{The fraction of recovered artificial stars as a function of magnitude for each filter. The detection limit was set to 5$\sigma$.}
\begin{tabular}{crrrr}
\tableline\tableline
Mag & F110W & F160W & F165M & F207M \\
17.0 & & & & 97\% \\
17.5 & & & & 97\%\\
18.0 & & & & 96\%\\
18.5 & & & & 77\%\\
19.0 & & & & 9\% \\
19.5 & & 98\% & 96\%  & 1\%\\
20.0 & 99\% & 98\% & 98\% &   \\
20.5 & 99\% & 94\% & 97\% & \\
21.0 & 97\% & 44\% & 51\% & \\
21.5 & 87\% & 5\% & 2\% & \\
22.0 & 66\% & 2\% & 1\% & \\
22.5 & 10\% & \\
23.0 & 2\% \\
\tableline
\tableline

\label{completeness}
\end{tabular}
\end{center}
\end{table}

\section{Results}
We present the basic results from the photometry. 
Objects are placed in the HR diagram in Section 3.1 based on the effective temperature  estimate derived  through the flux ratio between F165M and F160W and the dereddened J band magnitudes. 
In Section 3.2, we discuss the J--H versus J color--magnitude diagram and the H--K versus J--H color--color diagram. 
Throughout this section, we assume all but two very blue bright  stars are cluster members. 
The two blue stars have $J-H$ $\le$ 0.5 mag  and are assumed to be foreground objects. 
We have no direct estimate of the  number of  field stars expected towards Mon R2 which would contaminate our sample at the sensitivity of these observations.  
As an indirect test we have downloaded the 2MASS point sources located in a 2 degree region centered on (l=214.3 b=-15.4), where there coordinates for Mon R2 are (l=213.7,b=-12.6). 
The 2MASS H band luminosity function is a power--law down to the completeness limit of H$\sim$15 mag. 
Assuming the field star population would follow a power--law to faint magnitudes, we would expect $\sim$ 45 field stars within our field of view down to H=20.5 mag. 
However, the dust in the molecular cloud acts as a screen limiting the number of  background stars we will detect. 
\citet{carpenter} found the average extinction from the molecular cloud to be A$_\mathrm{V}=50$ mag in the central 45\arcsec\ region and A$_\mathrm{V}=33$ mag on average across the central 3\arcmin\ square region. 
If the field star luminosity function extrapolated from 2MASS is reddened by A$_\mathrm{V}=30$ mag, the surface density of field stars down to a limiting magnitude of H=20.5 mag is expected to be 1--2 objects per square arcminute. 
We have also estimated the  field star contamination using the synthetic Galactic model by \citet{robin}. 
The molecular cloud has been placed at 830pc and assumed to extinct background sources by A$_\mathrm{V}=30$ mag. 
The model predict that we should observe 10464 sources per square degree down to H=20.5 mag, where  6022 sources are predicted to be foreground sources. 
Thus, the model predicts 1.7 foreground objects and 1.2 background objects; the former consistent with our identification of 2 blue foreground objects,  the latter in agreement with the extrapolation from the 2MASS data.

\subsection{Effective temperatures and luminosities  for objects in Mon R2}
\begin{figure}
\epsscale{.65}
\plotone{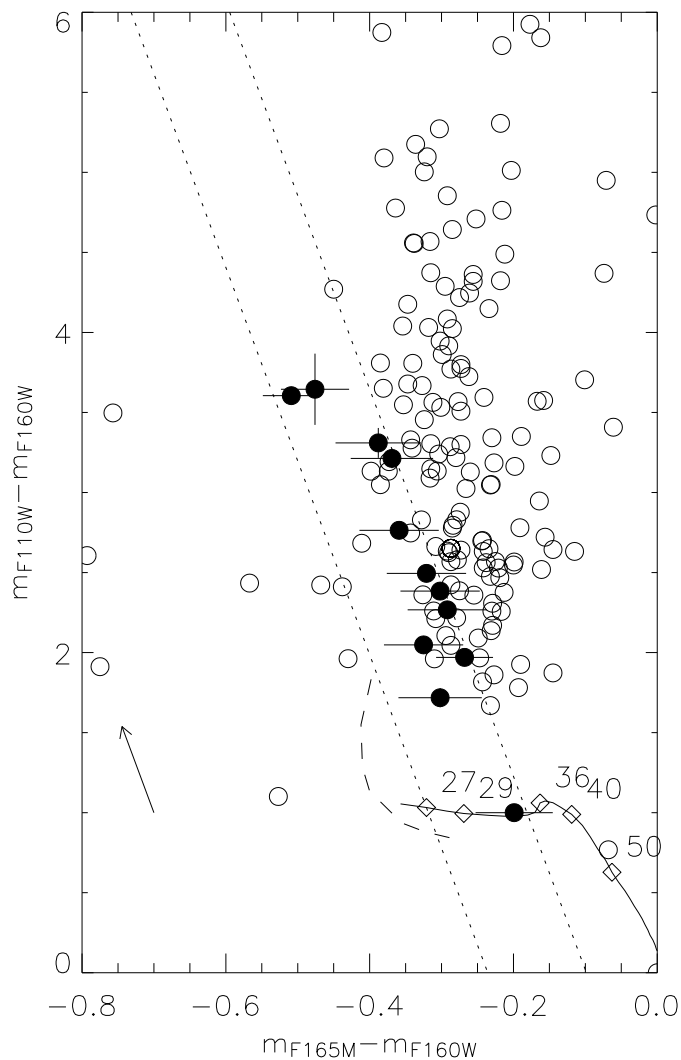}
\caption{The  $m_{F165M}-m_{F160W}$ versus $m_{F110W}-m_{F160W}$ color--color diagram. 
Overplotted are the DUSTY models by \citet{allard} for temperatures below 2500 K (dashed  line) and the NEXTGEN models by \citet{hauschildt} for temperatures above 2500 K (solid line). 
The arrow indicates the magnitude and slope of  A$_\mathrm{V}$=5 mag. 
The dotted lines indicate the region of the color--color diagram where a reliable temperature can be obtained if the photometric error in $m_{F165M}-m_{F160W}$ is less than $\sim$0.06 mag. 
Objects for which the effective temperature has been derived are shown as filled circles and the error bars are indicated for these objects. 
The effective temperatures at  positions on the two models are shown in units of 100 K and marked on the model with open diamonds. }
\label{temp_diag}
\end{figure}

A water vapor absorption feature is present in the H band in the spectra of stars and brown dwarfs with temperatures at or below 4000 K (see e.g. \citet{allard}). 
This feature is covered by the   F160W filter  and since the  F165M filter  measures the continuum, the color $m_{F165M}-m_{F160W}$ for an object cooler than $\sim$ 4000 K  can be used to estimate the effective temperature. 
We have compared the  $m_{F165M}-m_{F160W}$ colors,  with the synthetic colors predicted by the atmosphere models of \citet{hauschildt} and \citet{allard} for objects  above and below  2500 K, respectively. 
The $m_{F165M}-m_{F160W}$ versus $m_{F110W}-m_{F160W}$ color--color diagram is shown in Fig.~\ref{temp_diag}. 

\clearpage

\clearpage

 The reddening vector is parallel to the models for temperatures above $\sim$3800 K ($m_{F165M}-m_{F160W} >$ -0.1 mag) and no reliable temperature can be derived. 
There is a plateau from $\sim$3800 K and to $\sim$2700 K (spectral types M0 to M7.5) where an unique temperature can be obtained by projecting an observed point along the reddening vector until it intersects  the temperature  locus. 
The reddening vector  crosses  the models at two places for lower temperatures and a unique temperature cannot be obtained. 
We have estimated the effective temperature for the sources located between $\sim$3300 K and  $\sim$2700 K which have magnitude errors smaller than 0.06 mag. 
An uncertainty of 0.06 mag translates into an uncertainty of the temperature determined from the model of $\sim$10\%\ for this temperature range. 
We have adopted the models with a surface gravity of $\log g=4$, consistent with typical values of low mass PMS objects. 
We do not necessarily trust the magnitude of the extinction needed for the observations to intersect the temperature locus, due to uncertainties in the pseudo continua predicted by the models. 
Previous studies have shown the strength of the water  bands in the J and K bands only depend weakly on the surface gravity \citep{wilking,gorlova}. 
It could thus be expected the H band water vapor absorption feature is also weakly dependent of gravity. 
Indeed, adopting a model with $\log g=4.5$ only decreases the derived temperature by $\le$ 50 K. 
Decreasing the surface gravity to $\log g=3.5$ has a more complicated effect. 
Whereas the low temperature regime ($\le$ 3000 K) is unaffected, we would  underestimate the temperature by $\sim$300 K by incorrectly adopting the $\log g=4.0$ model for objects with temperatures between 3000 K and 3500 K. 
\citet{leggett2} compared the NextGen models to observed spectra of late type stars. 
They found the predicted  water bands to be deeper than observed for a given spectral type. 
The effect is that the models will predict a slightly too high temperature for a given waterband strength. 
\citet{najita00} and \citet{gorlova} find reasonable agreement between temperatures derived from comparison of water vapor indices from the models and spectral types. 
However, \citet{luhman2} found a systematic shift of spectral types relative to \citet{najita00} of 1--2 subtypes which they attributed to difference in the surface gravity between field dwarfs used to calibrate the index of pre--main sequence objects. 
Based on the model prediction of the temperature accuracy and the indirect evidence from other water bands,  we estimate our systematic uncertainties to be $\pm$300 K for temperatures derived using the water vapor index. 

\begin{figure}
\epsscale{.65}
\plotone{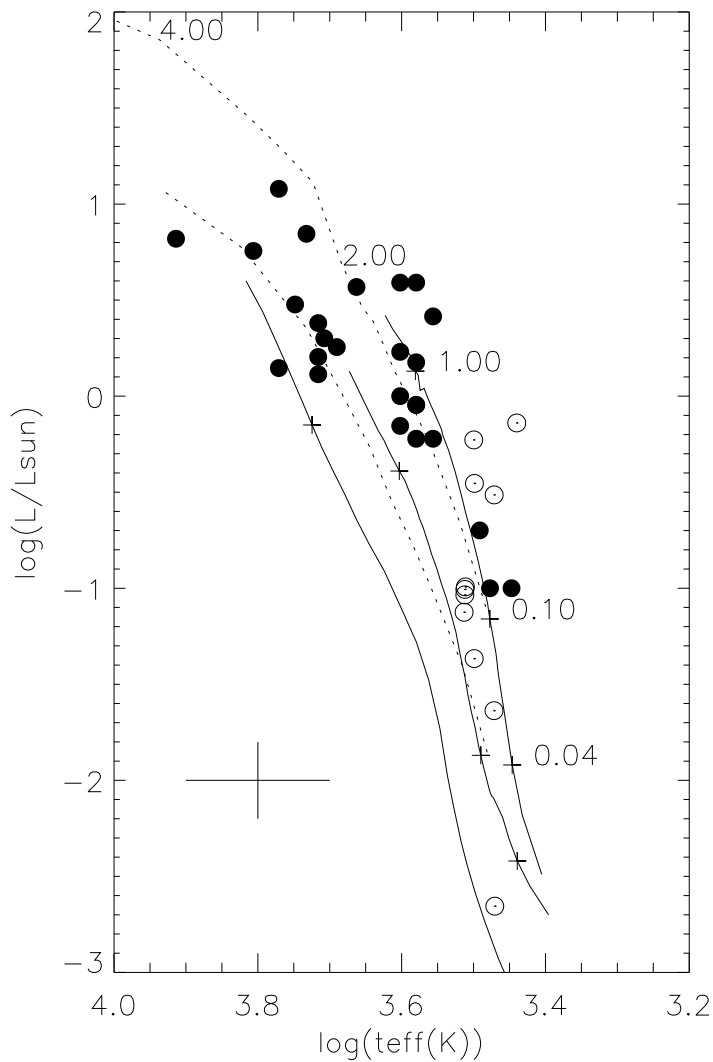}
\caption{The HR diagram based on  objects where the effective temperature and luminosity have been derived in this study using the $m_{F165M}-m_{F160W}$ color (open circles) and  objects from \citet{carpenter} (filled circles). 
Overplotted are the  1, 10, and 1000 Myr isochrones made from \citet{baraffe}  (solid lines)  and both \citet{pallastahler} 1 and 10 Myr isochrones (dashed lines). 
1 M$_\odot$, 0.1 M$_\odot$, and 0.04 M$_\odot$ have been marked on the Baraffe et al. isochrones with plus signs. 
 The \citet{pallastahler} isochrone covers the mass range 0.1--7 M$_\odot$. 
A typical error bar is shown.}
\label{HRD}
\end{figure}

Objects with a derived effective temperature in the range 2700--3300 K and with a photometric error  $m_{F165M}-m_{F160W} <$ 0.06 mag are marked with filled circles in Fig.~\ref{temp_diag}. 
We determined the bolometric luminosity  adopting formula 2 from \citet{gorlova}, which assumes $\mathrm{M_{bol}=4.64}$, and $\mathrm{BC_{V\odot}}=-0.19$. 
The dominant source of error in the bolometric luminosity stems from uncertainties in the derived extinction. 
An error per filter of 0.06 mag will translate into an error in the visual extinction A$_\mathrm{V}\sim$0.75 mag. Taking into account an error of $\sim$10\%\ in the bolometric correction, and the distance uncertainty, we  estimate  the error in the logarithm of the bolometric luminosity to be  0.2. 
The objects are then placed in the HR diagram, which is shown in Fig.~\ref{HRD}. 
We have not plotted source  83 in the HR diagram. 
Source 83 is a probable foreground source due to its very low  extinction. 
Source  6 also appears under--luminous. 
One possible reason for the under--luminosity can be the  extinction has been underestimated due to unresolved scattered light \citep[cf.][]{wilking}. 

The objects from \citet{carpenter} with effective temperature and luminosity estimates are shown as well. 
The 1, 10, and 1000 Myr \citet{baraffe} isochrones are shown. 
Since the \citet{baraffe} models only extends to 1.4 M$_\odot$ we have overplotted 1 Myr and 10 Myr  \citet{pallastahler} isochrones to cover masses up to 7 M$_\odot$. 
Table~\ref{waterbandsources} summarizes the parameters for the sources with the effective temperature derived from the water band method described above. 
\clearpage

\clearpage
Although a large fraction of the objects below 1 M$_\odot$ are located close to the 1 Myr isochrone, there appears to be a large scatter. 
Some the of scatter is due to errors in estimates of the effective temperatures and luminosities and  some of the scatter might be real. 
However, most of the objects that appear to be older than 1 Myr are more massive than 1 M$_\odot$, which is the maximum mass we attempt to constrain the IMF. 
We find that for the objects in the HR diagram with masses between 0.04 M$_\odot$ and 0.4 M$_\odot$ that 7 objects are 1 Myr or younger and 6 are older than 1 Myr. 
A median age of 1 Myr appears to be appropriate for the lower mass content in Mon R2. 

%

\begin{table}
\begin{center}
\caption{Physical parameters for the sources where we have estimated the effective temperature. }
\label{waterbandsources}

\begin{tabular}{cccccc}
\tableline\tableline
ID & A$_\mathrm{V}$\tablenotemark{a} & $\mathrm{T_{eff}}$ & $\mathrm{M_J}$ & excess\tablenotemark{b} & $\log L$ \\
\tableline
3 & 10.4 & 3250 & 5.29 & N & -1.00\\
6 & 1.4 & 2950 & 9.35 & Y & -2.66\\
13 & 9.4 &  3250 & 5.59 & N & -1.12\\
43 & 21.1 & 2750 & 2.93 & Y & -0.14\\
82 & 4.6 & 2950 & 4.00 &Y & -0.51 \\
83 & 0.0 & 3150 & 10.54 & Y &-3.11\\
105 & 17.7 & 3150 & 6.18 & Y & -1.37 \\ 
144 & 20.4 & 2950 & 6.81 & Y & -1.64 \\
147 & 14.2 & 3150 & 3.90 & N & -0.45\\
151 & 16.7 & 3250 & 5.37 & Y & -1.03\\
165 & 9.7 & 3150 & 3.33 & Y & -0.28 \\
171 & 6.5 & 3250 & 5.26 & N & -1.00 \\
\tableline
\tablenotetext{a}{Derived from the color--magnitude and color--color diagrams in Figs.~\ref{CMD_JH} and \ref{CCD} as described in Section 3.2.}
\tablenotetext{b}{Estimated from the color--color diagram in Fig.~\ref{CCD}.}

\end{tabular}
\end{center}
\end{table}

\clearpage

\subsection{Color-magnitude and color-color diagrams}
\begin{figure}
\epsscale{.8}
\plotone{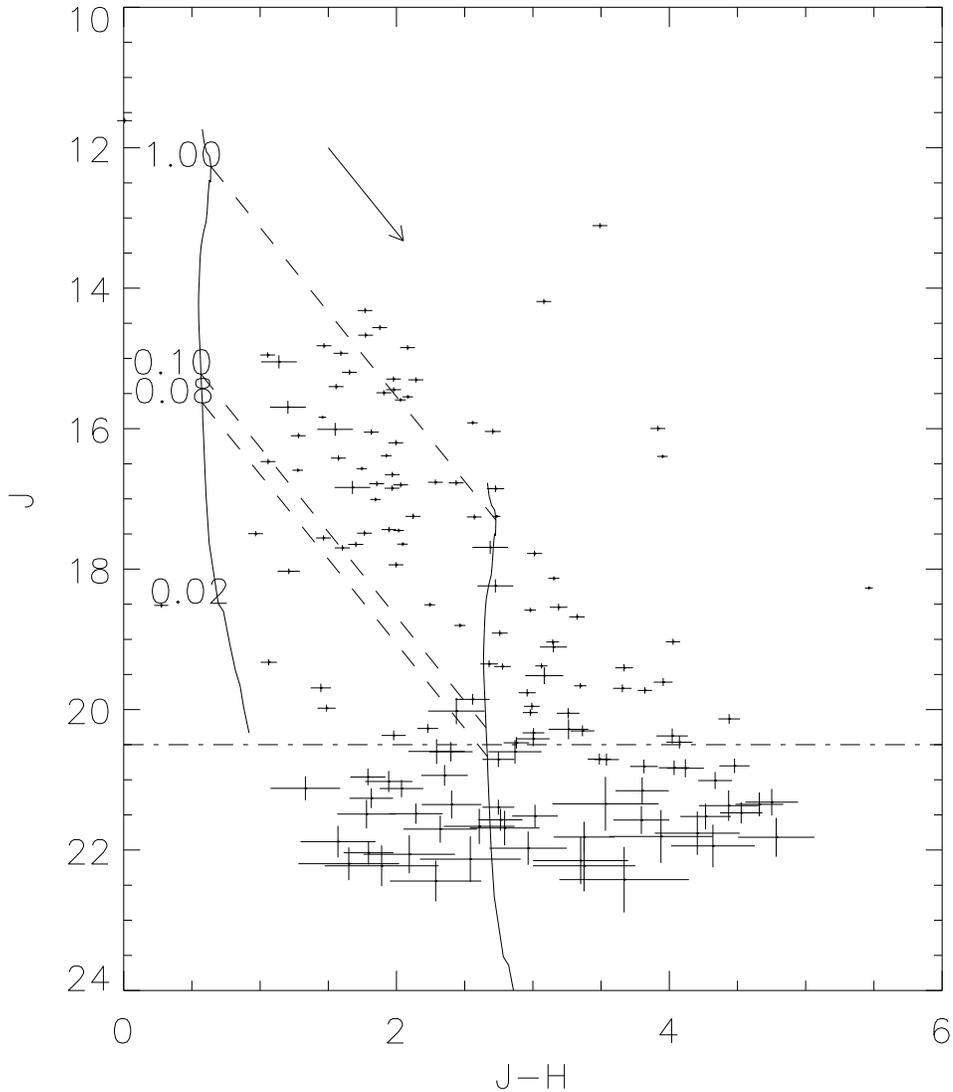}
\caption{The J--H versus J color--magnitude diagram for Mon R2 based on data from Table~2 converted into the CIT system. 
Overplotted as the solid lines are a  1 Myr isochrone  from the \citet{baraffe} models as discussed in the text, and the same isochrone reddened by A$_\mathrm{V}=19$ mag.  
The symbol sizes indicate photometric error, where the error in the transformation into the CIT system has been included.  
The horizontal line indicates the 90\% completeness limit of the observations. The  arrow illustrate the effect A$_\mathrm{V}=5$ mag extinction. 
Dashed lines are drawn between the unreddened and reddened isochrones at object masses of 1, 0.1, and 0.08 M$_\odot$.
The location of an unreddened 0.02 M$_\odot$ object is marked. }

\label{CMD_JH}
\end{figure}

We present in Fig.~\ref{CMD_JH} the J--H versus J color-magnitude diagram for the the inner 1\arcmin\ square of Mon R2. 

\clearpage

Also shown are a one Myr isochrone created from the \citet{baraffe} models, both unreddened and reddened by A$_\mathrm{V}=19$ mag and shifted with a distance modulus of 9.6 mag (830 pc). 
The Baraffe et al. models with $\mathrm{L_{mix}=H_P}$ are chosen since they cover the whole mass range where we attempt to constrain the IMF (0.02 M$_\odot$--1.0 M$_\odot$) and they are tied to the atmosphere models used in Section 3.1.

 In order to convert the effective temperatures and bolometric magnitudes provided by the models  to observables we have used the main sequence colors from  \citet{bessellbrett} and the temperature scales from \citet{schmidtkaler} and \citet{bessell} for objects earlier and later  than spectral type than K7, respectively. 
Data from \citet{dahn} have been used for  spectral types later than M6. 
The  bolometric corrections and colors  were interpolated using spline interpolation. 
Due to large scatter observed in the intrinsic colors and magnitudes for the late-type objects, a linear fit has been performed for objects later than M6. 

\begin{figure}
\epsscale{.8}
\plotone{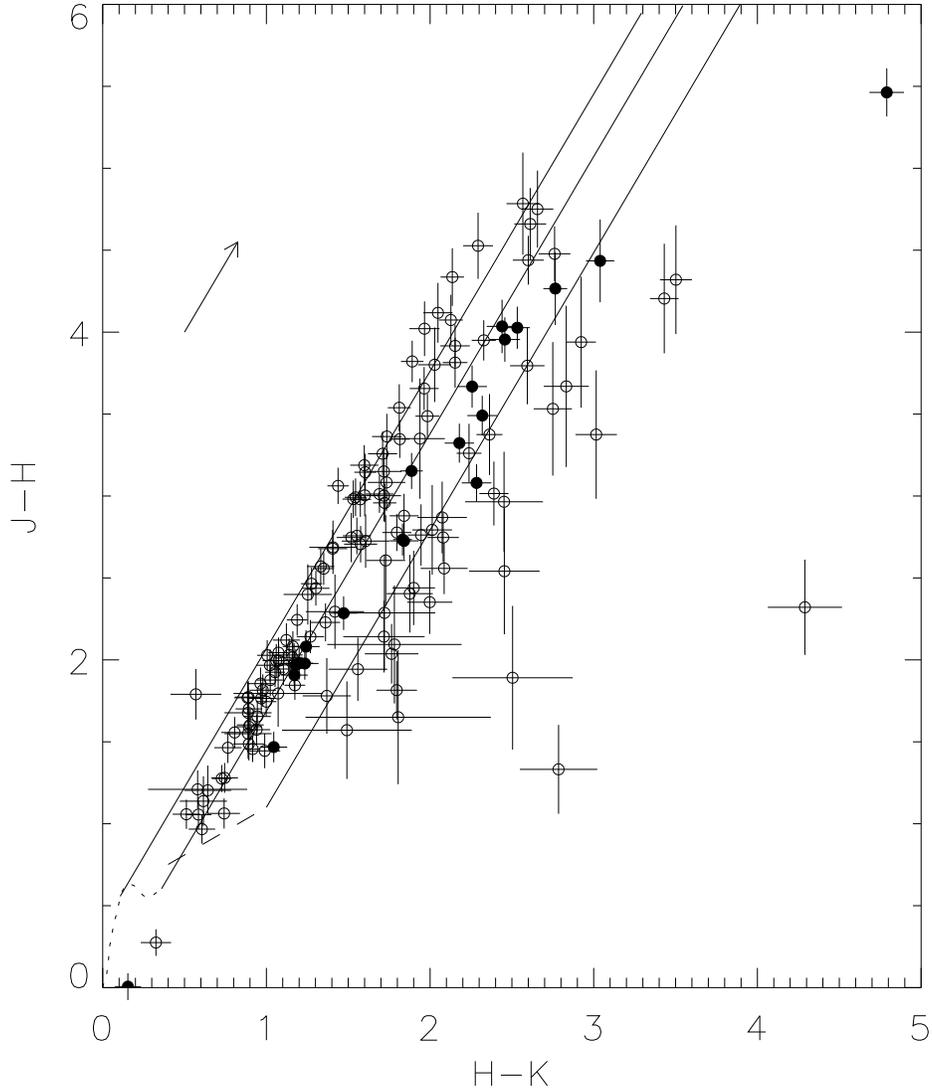}
\caption{The H--K versus J--H color-color diagram for Mon R2 based on data in Table~2 converted into the CIT system.  
The open circles  indicate stars not identified with any near-infrared excess whereas the filled circles indicate stars with near-infrared excess and an estimated spectral type earlier than M6 (see text). 
Shown as the dotted line are the colors from \citet{bessellbrett} for main sequence stars with spectral types B8 to M6. 
The length of the arrow illustrates the effect of A$_\mathrm{V}=5$ mag extinction. 
The  dashed line is the unreddened T-Tauri locus from \citet{meyer97}. 
Extended from the main sequence colors and  the classical T-Tauri slope are solid lines parallel to the reddening vector. }
\label{CCD}
\end{figure}

We expect the observed color of pre-main sequence objects  to be due to several effects. 
One is the intrinsic color of the object that depends on mass and age, which in turn fix $\mathrm{T_{eff}}$ and $\log(g)$.
Also, the  general interstellar extinction towards the object and  the possible presence of a circumstellar disk associated with the object. 
From the color-magnitude diagram it is not possible to distinguish between the two latter effects. 
However,  they can, at least partly,   be separated in the H--K versus J--H color-color diagram. 
Objects without  strong emission from a disk have colors resembling reddened  main sequence stars and  therefore populate a confined region in the color-color diagram. 
 Fig.~\ref{CCD} shows the color--color diagram for Mon R2 in the CIT system. 
Overplotted are the colors for main sequence stars from \citet{bessellbrett}, reddening vectors from stars with spectral type M6 and K7, and the dereddened classical T-Tauri locus derived  by \citet{meyer97}. 
Objects with an excess have a larger reddening  in H--K relative to what would be expected from their J--H color and are located to the right of the main sequence stars in the color-color diagram. 

\clearpage

\clearpage

A large fraction of the stars fall in the region expected for stars with no infrared excess and spectral type earlier than M6. 
The objects having larger than expected H--K relative to the reddening vector extending from the M6 main sequence star can  either be stars with spectral type earlier than M6 with an infrared excess or they can be objects of later spectral type  with or without infrared excess. 
The intrinsic colors for very late type objects (later than M6) almost overlap in the color--color diagram with the dereddened classical T-Tauri locus. 
Thus, we cannot tell for late spectral type objects whether they have a disk or not from the color-color diagram alone. 

The error adopted in $J-H$ by not being able to distinguish between an excess object and a late type object will be $\le 0.1$ mag, corresponding to A$_\mathrm{V}\le 1$ mag.  
We have therefore dereddened all objects in this part of the diagram to the T-Tauri locus in order to establish the intrinsic magnitudes of the objects. 
For stars with colors consistent with main sequence  stars they can in general be deredden to two different intersections with the main sequence colors. 
Therefore,  we have dereddened these stars  using the J--H color-magnitude diagram adopting the 1 Myr isochrone \citep{meyer96,carpenter,wilking}. 

\section{Analysis}
   In this section, we use the HST photometry to constrain the stellar and sub--stellar IMF in the
   central part of the Mon R2 cluster as well as re--visit the fraction of stellar mass objects with circumstellar disks. 
We begin by first defining an extinction limited sample that is complete down to 0.1 M$_\odot$. 
We use this sample to determine the fraction of stars with circumstellar disks and to search for variations in the extinction as a function of object mass. 
We then define three extinction limited samples complete for both stars and brown dwarfs in order to constrain the sub--stellar IMF of Mon R2 and compare these ratios with the field star IMF and other star    forming regions. 


\subsection{The circumstellar disk fraction for stars in Mon R2}
What is the fraction of stellar objects in our extinction limited sample showing evidence for a circumstellar disk as discussed above? 
We have constructed an extinction limited sample in order to obtain a representative sample  across the whole mass range considered. 
Without an extinction limited sample, we would underestimate the number of fainter,  lower mass sources relative to intrinsically more luminous objects of higher mass. 
The maximum extinction is determined such that the  sample considered will be complete down to spectral type M6, corresponding to a 0.1 M$_\odot$ star  for the 1 Myr isochrone. 
This corresponds to an extinction limit of A$_\mathrm{V}=19$ mag. 
We have excluded the two objects on the blue side of the isochrone in Fig.~\ref{CMD_JH} since these objects are probable foreground stars (star ID 61 and ID 83). 
An exclusion of these objects is equivalent to excluding objects with a derived extinction less than  A$_\mathrm{V}=1$ mag. 
The extinction limited sample (A$_\mathrm{V}=19$ mag) contains 43 objects in total. 
The selected sources are all located between the two upper  dashed lines in Fig.~\ref{CMD_JH}. 
In Section~4.4 we construct three further extinction limited samples in order to address the IMF into the brown dwarf regime.

For objects with a mass between  0.1 M$_\odot$ and 1.0 M$_\odot$, we find a disk fraction of $27\pm9\%$.  
For a wider area  but more shallow survey of Mon R2, \citet{carpenter} found a disk fraction of 48$\pm$8\% from JHK photometry. 
A disk fraction of $\sim$ 30\% is relatively low compared to what is found in other  star clusters of comparable age. 
For example, in the Orion Nebula Cluster (ONC), \citet{lada} found a disk fraction of $50\pm20\%$ using JHK photometry, \citep[cf.][]{hillenbrand_disks}, which within the error bars is consistent with Mon R2.  
\citet{haisch} found  a disk fraction of $21\pm5\%$ in IC 348 using JHK photometry.

\subsection{Constraints on the IMF in the center of Mon R2} 
With a dataset significantly deeper  than the previous study by \citet{carpenter}, we can constrain the IMF into the brown dwarf regime. 
The use of an extinction limited sample for IMF studies assumes the extinction distribution is independent of the objects mass. 
If this is not the case, a bias would  be introduced by preferentially excluding either high or low mass objects. 
To test if the extinction distribution is independent of object mass, we have divided our sample into objects between  0.2--1 M$_\odot$ and  between  0.03--0.2 M$_\odot$ for an A$_\mathrm{V}\le$ 10 mag extinction limited sample (see below). 
We have then performed a two sided   Kolmogorov--Smirnov (KS) test whether the extinction  distributions are consistent with having been drawn from the same parent population. 
The KS test returned a value of   0.49, from which we conclude there is no obvious difference in the extinction distribution as a function of mass. 

\begin{figure}
\epsscale{.85}
\plotone{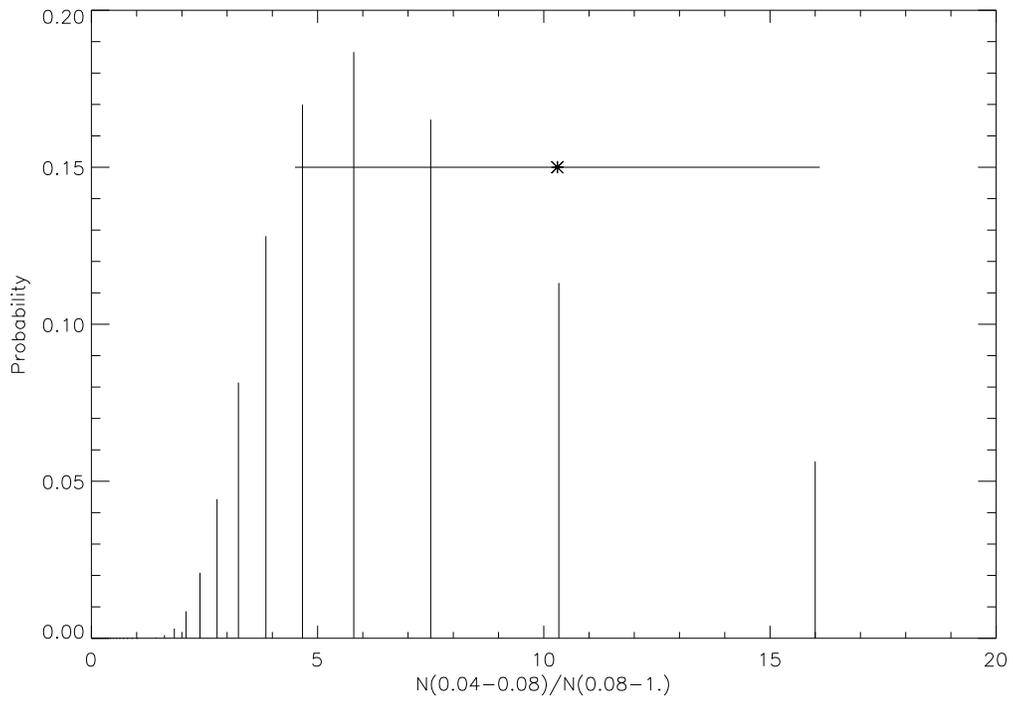}
\caption{A comparison of our derived R value with the R values obtained from a \citet{chabrierreview} field  IMF for a sample size of 34 objects. 
The predicted ratio distribution from the field IMF was determined used the binomial distribution.}
\label{kroupa_IMF}
\end{figure}

We have calculated  the ratio of stars between 0.08 and 1 M$_\odot$ divided by the number of objects between 0.04 and 0.08 M$_\odot$, $R$=N(0.08--1.0 M$_\odot$)/N(0.04--0.08 M$_\odot$), for an extinction limited sample A$_\mathrm{V} \le 13$ mag based on 34 objects. 
 The ratio is found to be  $R=10.3\pm5.8$ for a cluster age of 1 Myr, where the errors are derived assuming Poisson errors. 
We then compared the ratio with the similar predicted ratio from the \citet{chabrierreview} system field star IMF (dN/d$\log$M$\propto \exp{(-(logm-log0.22)^2/(2\cdot 0.57^2))}$). 
The distribution of predicted ratios is given by the binomial distribution and is shown in Fig.~\ref{kroupa_IMF}.
The height of each line shows the probability of a given ratio. 

\clearpage

\clearpage

\begin{figure}
\epsscale{.65}
\plotone{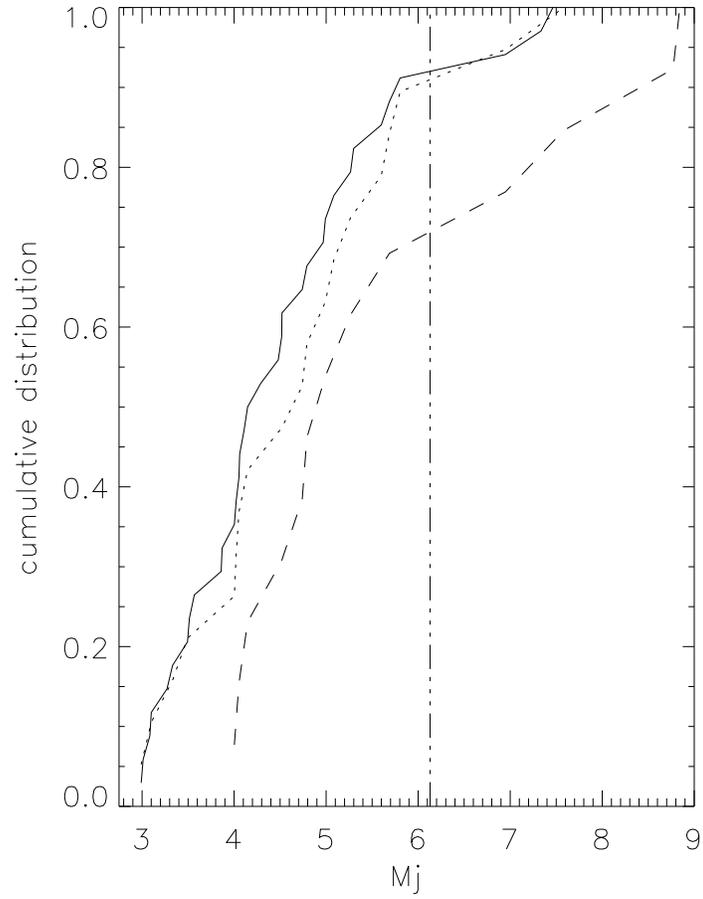}
\caption{The cumulative distributions of the absolute J band magnitudes for the objects in the  three extinction limited samples, A$_\mathrm{V}=13$ mag (solid line, 34 objects), A$_\mathrm{V}=10$ mag (dotted line, 19 objects), and A$_\mathrm{V}=7$ mag (dashed line, 13 objects). 
 The absolute magnitude of a 1 Myr 0.08 M$_\odot$ object \citep{baraffe} is marked with the vertical dot--dot--dashed line.}
\label{cummu}
\end{figure}

Although the ratio derived for Mon R2 is slightly higher than the peak of the distribution of ratios predicted by the \citet{chabrierreview} system IMF, the probability of obtaining the derived value or higher from the Chabrier system IMF is 19\%.
We have  constructed two additional extinction limited samples, one down to A$_\mathrm{V}=10$ mag and one to A$_\mathrm{V}=7$ mag. 
The two samples are complete down to 30 M$_\mathrm{jup}$, and 20 M$_\mathrm{jup}$, respectively. 
The upper mass for both samples is still 1 M$_\odot$. 
Although both samples extend to fainter absolute magnitudes than the A$_\mathrm{V}=13$ mag sample, the lower maximum extinction reduce the sample sizes significantly. 
Only 19  objects are included in the A$_\mathrm{V}=10$ mag sample and 13 in the A$_\mathrm{V}=7$ mag sample.
We have calculated the ratio of high--mass to low--mass objects for the two latter samples in a similar manner as above. 
We find the ratios $R^*$=N(0.08--1.0M$_\odot$)/N(0.03--0.08M$_\odot$)=8.5$\pm6.4$, and $R^{**}$=N(0.08--1.0M$_\odot$)/N(0.02--0.08M$_\odot$)=$2.2\pm1.3$. 
The similar expected ratios from the \citet{chabrierreview} system IMF are $R^*=4.2$, and $R^{**}=3.5$, respectively. 
It appears that the ratio of stars to brown dwarfs observed relative to the ratio expected from \citet{chabrierreview} decreases ($R$=10.3$\pm$5.8 vs 5.3;$R^*=8.5\pm6.4$ vs 4.3; $R^{**}=2.2\pm1.3$ vs 3.5) as a function of depth into the cloud. 
If confirmed, this might suggest that brown dwarfs and stars are not uniformly distributed in MonR2. 
However, the KS test did not indicate the distribution of extinction is different for the high and low mass samples, down to 0.03 M$_\odot$, so we do not believe these differences to be significant. 
All the observed ratios are consistent with the \citet{chabrierreview} IMF.  
The cumulative distributions of the absolute J band magnitudes for the three extinction limited samples are shown in Fig.~\ref{cummu}. 
The location of a  0.08 M$_\odot$ 1 Myr object from the \citet{baraffe} models is indicated.

\clearpage

\clearpage

\section{Discussion}
The derived ratios of high--mass to low--mass objects  can be compared with similar ratios derived from other young star clusters with similar spatial resolution. 
We will here focus on some of the best studied clusters, namely the ONC, IC348,  and Taurus and compare the values of $R, R^{*}$, and $R^{**}$ for those clusters with the values derived here. 
We will briefly discuss the selection criteria for the IMF samples for the different clusters and we compile the  ratios in Table~\ref{summarize}. 
 \begin{table}
\begin{center}
\caption{The derived ratios of high--mass to low--mass objects for Mon R2, the field star IMF and  other star forming regions.} 

\label{summarize}

\begin{tabular}{ccccccccc}
\tableline\tableline
& M$_\mathrm{low}$ &  Mon R2 & Field\tablenotemark{a}  & Field\tablenotemark{b} & IC348\tablenotemark{c} & Taurus\tablenotemark{d} & ONC\tablenotemark{e} & ONC\tablenotemark{f}\\
\tableline
$R$ &0.04 & 10.3$\pm$5.8 & 2.9 & 5.3 & 16.8$\pm$5.8 & 9.6$\pm$3.2 & 6.4 & 5.5$\pm$0.8\\
$R^*$& 0.03 & 8.5$\pm$6.4 & 2.2 & 4.2 & 11.6$\pm3.4$ & 6.9$\pm$2.0 & 4.7 & 4.3$\pm$0.6\\
$R^{**}$ & 0.02 & 2.2$\pm$1.3 & 1.7 & 3.5 &    & 5.4$\pm$1.4 & 3.7 & 3.0$\pm$0.6\\
\tablenotetext{a}{\citet{kro02}, single star IMF}
\tablenotetext{b}{\citet{chabrierreview}, system IMF}
\tablenotetext{c}{\citet{luhman2}}
\tablenotetext{d}{\citet{luhman}}
\tablenotetext{e}{\citet{muench}}
\tablenotetext{f}{\citet{carphil}}
\end{tabular}
\end{center}
\end{table}

As the closest site of massive star formation, the ONC has been observed in some detail. 
The IMF has been estimated through deep near--infrared imaging \citep{carphil,muench}. 
Using H and K band observations, \citet{carphil} derived the IMF down to 0.02 M$_\odot$ for A$_\mathrm{V}\le 10$ mag. 
From their histogram we have derived the  ratios for the ONC where we find $R=5.5\pm0.8$, $R^{*}=4.3\pm0.6$, and $R^{**}=3.0\pm0.6$, respectively.  
In another imaging study, the IMF of the ONC was constrained through model fits to the K band luminosity function by \citet{muench} down to $\sim$10 M$_\mathrm{jup}$. 
The ratio of stars to brown dwarfs was found by integrating the functional fit to the IMF presented by \citet{muench}. 
From the functional fit, we find $R=6.4$, $R^*=4.7$, and $R^{**}=3.7$, respectively consistent both with result by \citet{carphil} and our ratios for Mon R2.

Spectroscopic observations of objects in the ONC by \citet{slesnick} indicate the ratio of high mass to low mass objects  might be  higher than found from  imaging alone.
A ratio of R$'$=N(0.08-0.4 M$_\odot$)/N(0.03-0.08 M$_\odot)=$4.5$\pm0.7$ is found from the spectroscopic survey. 
It should be kept in mind that correction factors of $\sim$2 have been applied to the number counts as discussed in \citet{slesnick}.  
The comparable ratio for Mon R2 is  4.5$\pm$3.5 which is in excellent agreement with the \citet{slesnick} result. 
For comparison, the similar ratio from the \citet{muench} IMF is 4.1. 

As one of the nearest low--mass star forming regions to the Sun, Taurus has been investigated in detail. 
\citet{luhman} has presented the IMF down to 0.015 M$_\odot$ for an extinction limited sample A$_\mathrm{V}\le 4$ mag. 
We have calculated the ratio of high to low--mass stars from the data presented by  \citet{luhman}. 
We find the derived IMFs for Taurus and Mon R2 to be consistent within the uncertainties for all three ratios.

The ratios for Taurus might be suspect for two reasons. 
\citet{luhman} emphasized the sample is incomplete for the mass interval 0.3--0.6 M$_\odot$ due to saturation, resulting in the derived ratios being lower limits. 
Also, the surveyed region by \citet{luhman} only covered parts of Taurus (4 square degrees), preferentially centered on the dense filaments.  
A larger area survey (28 square degrees) has found relatively more brown dwarfs \citep{guieu}, although the latter survey  only extended down to 0.03 M$_\odot$, for an extinction limited sample A$_\mathrm{V}\le 4$ mag.  
They  concluded the ratio of brown dwarfs to star in Taurus and the ONC are similar within the statistical uncertainties. 
Although the ratio derived by \citet{guieu} extended to higher masses relative to this paper, we can still estimate the effect of their updated number of brown dwarfs in Taurus. 
The newly detected brown dwarfs in the larger surveyed region by \citet{guieu} reduced the derived ratios by $\sim$ 25\%\ relative to \citet{luhman}. 
Correcting the Taurus results quoted in Table~\ref{summarize} downwards by 25\%\ only improves the agreement with the ratios for Mon R2. 

The low-mass stellar content of IC348 has been investigated  recently by \citet{luhman2}. 
The IMF was derived down to 30 M$_\mathrm{jup}$, using an extinction limited sample A$_\mathrm{V}\le 4$ mag. 
The objects presented by \citet{luhman2} have been  spectroscopically confirmed as members. 
The spatial resolution in the imaging survey was 0\farcs6. Scaled to the distance of Mon R2, this is a similar physical  resolution as our F207M band image. 
Based on the data published in \citet{luhman}, we have derived the ratios $R=16.8\pm5.8$ and $R^*=11.6\pm3.3$. 
The ratios for IC 348 appears to be relatively high but still consistent within 2$\sigma$ to the results for Mon R2. 
The IMF has been probed in IC 348 down to 0.015 M$_\odot$ utilizing HST/NICMOS imaging \citep{najita00}. 
Using the the strength of the 1.9$\mu$m water band they determined the spectral type of objects in the mass interval 0.015--0.7 M$_\odot$. 
We have calculated the modified ratio of high--mass to low--mass objects using objects between 0.25--0.7 M$_\odot$ as the high--mass objects and 0.02--0.25 M$_\odot$ as the low--mass objects. 
The ratio derived from the histogram in \citet{najita00} is 0.4, whereas the similar ratio for Mon R2 is 0.3$\pm0.2$ for an extinction limited sample, A$_\mathrm{V}\le 7$ mag. 


One issue that should be explicitly  taken into account in these comparisons is the role of binaries. 
It is well known that a large fraction of stars reside in multiple systems and 
due to our finite angular resolution, we can only resolve a certain fraction of the binary systems in each cluster. 
The resolution in the F207M band  with HST/NICMOS  at a distance of 830 pc, resolves all systems more widely separated than $\sim$160 AU. 
Both the ONC imaging surveys have a similar physical resolution with the  cluster  at 430 pc. 
\citet{luhman} considered all objects with a separation smaller than 2\arcsec\ in Taurus as single objects. 
For a distance of 140 pc, this corresponds to a separation of 280 AU. 
On the other hand, the HST/NICMOS imaging of IC 348 by \citet{najita00} has twice the physical resolution as this study, resolving binaries with separation larger than $\sim$ 75 AU. 
Since the physical angular resolution is similar within a factor of $\sim$2 for most of the clusters compared, we do not expect unresolved binaries to effect the ratios differently for the clusters discussed here. 
However, the field IMF by \cite{kro02} has been constructed from individual objects. 
The majority of binaries in young open clusters have separations smaller than $\sim$ 100 AU \citep[][cf. \citet{DM91}]{patience} and would be unresolvable in our observations. 
The quantitative change of the IMF due to unresolved binaries  depends on the details of the binary frequency versus separation  and the distribution of mass ratios, neither of which are well known, especially for  low masses. 
However, there is some indication the binary frequency is lower for cool objects \citep{burgasser}. 
\citet{chabrierreview} have presented the system IMF assuming no unresolved binaries derived from the local 8 pc solar neighborhood field star sample, which is used in Table~\ref{summarize}. 

The derived ratios for Mon R2 appears consistent with the similar ratios for the ONC, IC 348,  and Taurus. 
They are also consistent with the ratios derived for both the \citet{kro02} and \citet{chabrierreview} IMF. 
Thus, despite more than an order of magnitude difference in total cluster mass, it appears the IMF down to 0.02 M$_\odot$ is similar. 
{\it So far, there is little evidence for variations in the sub--stellar IMF at least down to 20 M$_{jup}$.}
It remains to be seen whether or not the ensemble of the observations can distinguish between a flat or a falling IMF between 0.02--0.08 M$_\odot$ \citep[cf.][]{allen}.

Several theoretical considerations would predict  a  deficit of brown dwarfs in Taurus relative to more massive regions. 
\citet{goodwin} suggest the IMF in a region like Taurus should have relatively fewer brown dwarfs due to the narrow distribution of core masses in Taurus relative to regions like the ONC. 
An alternative explanation has been presented by \citet{batebonnell}.  
They suggested the lower density in the Taurus molecular cores results in a higher Jeans mass which in turns results in a higher average mass. 
\citet{goodwin2} indicate through numerical simulations that the peak of the IMF will shift to lower masses as the degree of turbulence increases.

The  increase of the relative number of brown dwarfs when the minimum mass of the sample is decreased is curious. 
Since the extinction limit decreases as the limiting mass is extended to lower masses, one possibility is that the more massive stars are observed preferentially deeper within the molecular cloud. 
However, we found no evidence for  variation in the extinction as a function of object mass down to 0.03 M$_\odot$. 
One possibility, although speculative, is that the brown dwarfs have been ejected from small N body systems as proposed by e.g. \citet{reipurth}. 
We would then detect the brown dwarfs ejected towards us (preferentially with lower extinction) even though the parent system containing the star would be located deeper in the cloud. 
Since we are only probing the surface of the cluster in the most shallow extinction limited samples, such a scenario would explain the observed trend. 
Deeper analysis of the dynamical evolution (velocities and spatial distribution) of cluster members as a function of mass is needed. 

\section{Conclusions}
We have presented the results from HST/NICMOS2 F110W, F160W, F165M, and F207M band imaging of  the inner 1$\arcmin\times1\arcmin$ of the  embedded cluster associated with  Mon R2. 
Our results are as follows:
\begin{itemize}
\item{The effective temperature has been estimated for a small set of stars in the temperature range $2700-3300$ K based on a water vapor index and these objects have been placed in the HR diagram. 
A 1 Myr \citet{baraffe} isochrone is consistent with the lower mass objects placed in the HR diagram in agreement with \citet{carpenter}.}
\item{We find that for stars with spectral type M6 or earlier (0.1--1M$_\odot$) a disk fraction of $27\pm9$\% based on an extinction limited sample of 43 stars A$_\mathrm{V} \le 19$ mag. }
\item{We created three extinction limited samples complete for A$_\mathrm{V}=13$, 10, and 7 mag, and containing 34, 19, and 13 objects, respectively.  
We have calculated the  three ratios of low mass stars to brown dwarfs $R=N(0.08-1.0 \mathrm{M}_{\odot})/(0.04-0.08 \mathrm{M}_\odot)=10.3\pm5.8$, $R^{*}=N(0.08-1.0\mathrm{M}_{\odot})/N(0.03-0.08 \mathrm{M}_{\odot})=8.5\pm6.4$ and $R^{**}=N(0.08-1.0 \mathrm{M}_{\odot})/N(0.02-0.08 \mathrm{M}_{\odot})=2.2\pm1.3$. }
\item{The derived ratios are consistent with the similar ratios for  Taurus, IC 348, the ONC, and the system field IMF of \citet{chabrierreview}. 
There is thus no compelling evidence for variations in the relative brown dwarf content between Mon R2 and other nearby star forming regions. }
\end{itemize}






\acknowledgments
This work was supported by a Cottrell Scholar's Award to MRM from the Research Corporation and NASA grant HST13-9846. 
We would like to thank Catherine Slesnick and Lynne Hillenbrand for providing the data from their spectroscopic survey of the ONC. 
We thank Steve Strom for discussions during the early stages of the project, France Allard for assistance with the models, Erick Young and Hua Chen for help in preparations of the observations, Angela Cotera for advice concerning the image processing and Kevin Luhman for assistance in calculating the mass ratios for IC 348 and Taurus. 
We also thank the referee August Muench for comments and suggestions that improved and clarified the paper.



Facilities: \facility{This paper is based on observations made with the NASA/ESA { \it Hubble Space Telescope}, operated by the Space Telescope Science Institute, which is operated by the Association of Universities for Research in Astronomy, Inc., under NASA contract NAS5-26555.}.

\clearpage

\end{document}